\documentclass[useAMS,usenatbib]{mn2e}

\usepackage{graphicx}
\usepackage{caption}
\usepackage{times}%
\usepackage{natbib}
\usepackage{graphics}
\usepackage{epsfig}
\usepackage{array}
\usepackage{stfloats}
\usepackage{fixltx2e}
\usepackage{amsmath}
\usepackage[a4paper,colorlinks=true,pdfstartview=FitV,linkcolor=red,citecolor=blue,urlcolor=magenta]{hyperref}
\usepackage{float}

\newcommand{\lesssim}{\la} 

\newcommand{\ie}{{i.e.}}

\newcommand{\gsim}{\,\lower2truept\hbox{${>\atop\hbox{\raise4truept\hbox{$\sim$}}}$}\,}

\def\ie{{\rm i.e.$\,$}}

\newcommand{\be}{\begin{equation}}
\newcommand{\ee}{\end{equation}}
\newcommand{\bea}{\begin{eqnarray}}
\newcommand{\eea}{\end{eqnarray}}

\def\ltsima{$\; \buildrel < \over \sim \;$}
\def\simlt{\lower.5ex\hbox{\ltsima}}
\def\gtsima{$\; \buildrel > \over \sim \;$}
\def\simgt{\lower.5ex\hbox{\gtsima}}

\title[Cosmic Degeneracies I: $f(R)$ and massive neutrinos]{Cosmic Degeneracies I: Joint N-body Simulations of Modified Gravity and Massive Neutrinos}
\author[M. Baldi et al.]{\parbox{\textwidth}{Marco Baldi$^{1,2,3}$, Francisco Villaescusa-Navarro$^{4}$, Matteo Viel$^{4,5}$, Ewald Puchwein$^{6,7}$, \\Volker Springel$^{7,8}$, Lauro Moscardini$^{1,2,3}$}
\\
\\$^{1}$Dipartimento di Fisica e Astronomia, Alma Mater Studiorum Universit\`a di Bologna, viale Berti Pichat, 6/2, I-40127 Bologna, Italy;
\\$^{2}$INAF - Osservatorio Astronomico di Bologna, via Ranzani 1, I-40127 Bologna, Italy;
\\$^{3}$INFN - Sezione di Bologna, viale Berti Pichat 6/2, I-40127 Bologna, Italy;
\\$^{4}$INAF - Osservatorio Astronomico di Trieste, Via Tiepolo 11, I-34143, Trieste, Italy;
\\$^{5}$INFN/National Institute for Nuclear Physics, Via Valerio 2, I-34127 Trieste, Italy;
\\$^{6}$Institute of Astronomy and Kavli Institute for Cosmology, University of Cambridge, Madingley Road, Cambridge CB3 0HA, UK;
\\$^{7}$Heidelberger Institut f\"ur Theoretische Studien (HITS), Schloss-Wolfsbrunnenweg 35, 69118 Heidelberg, Germany;
\\$^{8}$Zentrum f\"ur Astronomie der Universit\"at Heidelberg, ARI, M\"onchhofstrasse 12-14, 69120 Heidelberg, Germany.}

\begin{document}
%\date{Accepted ???. Received ???; in original form }
\pagerange{\pageref{firstpage}--\pageref{lastpage}} \pubyear{2011}
\maketitle
\label{firstpage}
\begin{abstract}	
\\
We present the first suite of cosmological N-body simulations that simultaneously include the effects of two different and theoretically independent
extensions of the standard $\Lambda $CDM cosmological scenario -- namely an $f(R)$ theory of Modified Gravity (MG) and a cosmological
background of massive neutrinos -- with the aim to investigate { their} possible observational degeneracies. We focus on three basic statistics of the large-scale matter distribution, more specifically the nonlinear matter power spectrum, the halo mass function, and the halo bias.
%, for which we determine the deviation with respect to the fiducial $\Lambda $CDM cosmology in the context of both separate and combined simulations of $f(R)$ MG and massive neutrinos scenarios. 
Our results show that while these two extended models separately determine very prominent 
and potentially detectable 
features in all the three statistics, when we allow them to be simultaneously at work { these} features are strongly suppressed.
%, resulting in much weaker deviations from the standard {model's} predictions. 
%Remarkably, { a specific} combination of the characteristic parameters of the two models { provides an effective cancellation} in all the three statistics at $z=0$, and { this}
%broad degeneracy is only weakly broken when the redshift evolution of the different observables { is considered}. 
In particular, when an $f(R)$ gravity model with $f_{R0}=-1\times 10^{-4}$ is combined with a total neutrino mass of $\Sigma _{i}m_{\nu _{i}}=0.4$ eV, the resulting matter power spectrum, halo mass function, and bias at $z=0$ are found to be consistent with the standard model's predictions at the $\lesssim 10\%$, $\lesssim 20\%$, and $\lesssim 5\%$ accuracy levels, respectively.
Therefore, our results imply an intrinsic theoretical limit to the effective discriminating power of present and future observational data sets with respect to these widely considered extensions of the standard cosmological scenario. 
%{ in the absence of independent measurements of the neutrino masses from laboratory experiments, { even though the high-redshift evolution might still allow to partially break the degeneracy.}}
\end{abstract}


%*****************************************************************************

\section{Introduction}
\label{i}

The primary scientific goal of a wide range of present and future observational initiatives
in the field of cosmology -- such as e.g. { BOSS \citep[][]{Ahn_etal_2013}, }
PanStarrs
\citep[][]{PanStarrs}, HETDEX \citep[][]{HETDEX}, DES \citep[][]{DES},
LSST \citep[][]{LSST} and Euclid\footnote{www.euclid-ec.org}
\citep[][]{EUCLID-r} -- consists in combining different probes of the geometric and dynamical
evolution of the Universe to unveil possible deviations from the expected behaviour of the 
standard $\Lambda $CDM cosmological scenario. The latter relies on the assumption of a 
cosmological constant $\Lambda $ as the source of the observed accelerated expansion of the Universe 
\citep[][]{Riess_etal_1998,Perlmutter_etal_1999,Schmidt_etal_1998} and on the existence of some yet undetected
Cold Dark Matter (CDM) particle with negligible non-gravitational interactions that could enhance the growth of cosmic structures from the
tiny density perturbations observed in the primordial Universe through detailed Cosmic Microwave Background (CMB) 
observations \citep[see e.g.][]{Smoot_etal_1992,wmap9,Planck_016} to the highly structured Universe that we observe today { \citep[see e.g.][]{SDSS-7}.} 

Although the standard $\Lambda $CDM cosmological model { has so far been found} to be consistent with observational data
at ever increasing levels of accuracy \citep[see e.g.][]{Blake_etal_2011,Planck_016}, its theoretical foundations remain
poorly motivated in the absence of a clear understanding of the physical origin of the cosmological constant $\Lambda $ and of
a direct (or indirect) detection of the elementary particle associated with CDM. { This} lack of a firm theoretical and observational { basis}
for the standard $\Lambda $CDM picture has motivated the exploration of a large number of alternative scenarios both concerning the { origin} of the accelerated expansion and the nature of the CDM cosmic field. 

The former range of extended models {includes} on one hand 
generalised Dark Energy (DE) scenarios where the cosmic acceleration is driven by a classical
scalar field \citep[see e.g.][]{Wetterich_1988,Ratra_Peebles_1988,kessence,Caldwell_2002,Feng_Wang_Zhang_2005} possibly characterised by non-trivial clustering \citep[see e.g.][]{Creminelli_etal_2009,Sefusatti_Vernizzi_2011,Batista_Pace_2013} or interaction
\citep[see e.g.][]{Wetterich_1995,Amendola_2000,Farrar2004,Amendola_Baldi_Wetterich_2008,Baldi_2011a} properties, and on the other hand several possible Modified Gravity (MG) models including {\em Scalar-Tensor
  theories} of gravity \citep[as e.g. $f(R)$ models,
][]{Buchdahl_1970,Starobinsky_1980,Hu_Sawicki_2007,Sotiriou_Faraoni_2010}, the
{\em DGP} scenario \citep[][]{Dvali_Gabadadze_Porrati_2000}, or the
{\em Galileon} model \citep[][]{Nicolis_Rattazzi_Trincherini_2009}. In the present work we will focus on { the} second class of extensions of the standard cosmology, and in particular on the widely investigated parameterisation of $f(R)$ theories
of gravity proposed by \citet{Hu_Sawicki_2007}.

The latter range of models, instead, { includes} various extensions of the CDM paradigm characterised by different assumptions
on the nature of the fundamental dark matter particle, both concerning its phase-space distribution -- as for the case of Warm Dark Matter (WDM) models \citep[see e.g.][]{Bode_Ostriker_Turok_2001,Viel_etal_2013} or mixed dark matter scenarios \citep[][]{Maccio_etal_2012} -- and its 
interaction properties \citep[see e.g.][]{Loeb_Weiner_2011,Baldi_2013}. While the fundamental composition of the dark matter fraction of the total cosmic
energy budget has very little impact on the background expansion history of the universe, it can significantly affect the growth of density perturbations both in the linear and in the nonlinear regimes. In particular, one of the most significant extensions of the standard  cosmological scenario for what concerns { the} structure formation processes amounts to dropping the assumption of massless neutrinos that is commonly adopted in both analytical and numerical investigations of the late-time universe. The discovery of the neutrino oscillation phenomena \citep[][]{Cleveland_etal_1998} has
revealed in an unambiguous way that at least two of the three neutrino families are massive, such that a fraction of the total matter density of the
Universe {\em must} be associated { with} the cosmic neutrino background. In this respect, the inclusion of massive neutrinos into any cosmological
scenario should no longer be regarded as one (amongst many) possible extension of the basic standard model into the realm of exotic physics, but rather as a necessary ingredient in order to faithfully reproduce reality. As such, any realistic cosmological modelling -- both for the standard cosmological constant $\Lambda $ and for alternative DE or MG scenarios -- should no longer avoid to properly take into account the possibility that part of the { dark matter} energy density be made of non-relativistic massive neutrinos. 

In recent years, significant progress has been made in including the effects of massive neutrinos into cosmological N-body codes employed to study structure formation processes from the linear to the highly nonlinear regime in the context of standard $\Lambda $CDM cosmologies {\citep[see e.g.][]{Brandbyge_etal_2008,Brandbyge_Hannestad_2009,Brandbyge_Hannestad_2010,Viel_Haehnelt_Springel_2010,Agarwal_Feldman_2011,Bird_Viel_Haehnelt_2012,Wagner_Verde_Jimenez_2012}.}
Here, by ``standard $\Lambda $CDM cosmologies" we refer to models where the accelerated cosmic expansion is driven by a cosmological constant and where a fixed { total matter} density is made up by a fraction of standard CDM particles and by a fraction of massive neutrinos.
Such works have highlighted a number of effects that appreciably modify observable quantities \citep[as e.g. the nonlinear matter power spectrum, the abundance of collapsed halos as a function of their mass, the pattern of redshift-space distortions, { or the clustering properties of dark matter halos}, see e.g.][respectively]{Viel_Haehnelt_Springel_2010,Castorina_etal_2013,Marulli_etal_2011,Villaescusa-Navarro_etal_2013} in massive neutrinos cosmologies as compared to their massless neutrinos counterparts, thereby possibly biasing the inference of cosmological parameters from these observables. 
Similarly, significant progress has been made in implementing alternative DE and MG models in N-body algorithms \citep[see e.g.][for a recent review]{Baldi_2012b}, which have allowed self-consistent cosmological simulations in the context of a wide range of such non-standard scenarios for the accelerated cosmic expansion.
However, no attempt has been made so far to combine such two efforts and investigate the effects of massive neutrinos on the formation and evolution of linear and nonlinear cosmic structures in the context of alternative DE and MG models. \\

In the present paper -- which is the first in a series of works aimed at studying the intrinsic observational degeneracies  between different and independent extensions of the standard $\Lambda $CDM scenario -- we attempt for the first time to investigate the joint effects of $f(R)$ MG theories and massive neutrinos in the nonlinear regime of structure formation by means of a specifically designed N-body code that is capable of simultaneously including both { these} additional  ingredients in its integration scheme. 
The latter is a modified version of the widely used TreePM N-body code {\small GADGET} \citep[][]{gadget-2} which has been obtained by combining the {\small MG-GADGET} implementation of $f(R)$ theories recently developed by \citet{Puchwein_Baldi_Springel_2013}
with the massive neutrinos module by \citet{Viel_Haehnelt_Springel_2010}. 

With such code { in} hand, we have performed for the first time a series of combined simulations for one specific realisation of $f(R)$ MG  with different amounts of massive neutrinos, and compared the outcomes of our runs in terms of some basic statistics of the { Large-Scale Structures (LSS)} distribution to the same set of neutrino masses acting within a standard $\Lambda $CDM cosmology. { This work therefore extends the earlier studies of \citet{He_2013} and \citet{Motohashi_etal_2013} to the fully nonlinear regime of structure formation.} Our results highlight a very strong degeneracy between $f(R)$ MG models and massive neutrinos in all the observables that we have considered, such that a suitable combination of $f(R)$ parameters and neutrino masses might be hardly distinguishable from the standard cosmological scenario even when the two separate effects would result in a very significant deviation from the reference cosmology. { This} degeneracy should be properly taken into account when assessing the effective discriminating power of present and future observational surveys with respect to
individual extensions of the standard cosmological model. 

The present paper is organised as follows. In Section~\ref{foR} we introduce the $f(R)$ MG models considered in our numerical simulations. In Section~\ref{intro_neutrinos} we review the basics of massive neutrinos and their role in cosmological structure formation. In Section~\ref{sims} we describe the numerical setup of our cosmological simulations and of the initial conditions generation. In Section~\ref{res} we present our results on the nonlinear matter power spectrum, the halo mass function, and the halo bias. Finally, in Section~\ref{concl} we draw our conclusions.

\begin{table*}
\begin{tabular}{lcccccc}
\hline
Run & Theory of Gravity  &  
$\Sigma _{i}m_{\nu _{i}} $ [eV] &
$\Omega _{\rm CDM}$ &
$\Omega _{\nu }$ &
$M^{p}_{\rm CDM} [M_{\odot }/h]$ &
$M^{p}_{\nu } [M_{\odot }/h]$\\
\hline
GR-nu0 & GR & $0$ & $0.3175$ & $0$ &$6.6\times 10^{11}$ & $0$ \\
GR-nu0.2 & GR & $0.2$ & $0.3127$ & $0.0048$ & $6.47\times 10^{11}$ & $9.86\times 10^{9}$  \\
GR-nu0.4 & GR & $0.4$ & $0.308$ & $0.0095$ & $6.4\times 10^{11}$ & $1.97\times 10^{10}$\\
GR-nu0.6 & GR & $0.6$ & $0.3032$ & $0.0143$ & $6.27\times 10^{11}$ & $2.96\times 10^{10}$ \\
fR-nu0 & $\bar{f}_{R0} = -1\times 10^{-4}$ & $0$ & $0.3175$ & $0$ & $6.6\times 10^{11}$ & $0$ \\
fR-nu0.2 & $\bar{f}_{R0} = -1\times 10^{-4}$ & $0.2$ & $0.3127$ & $0.0048$ & $6.47\times 10^{11}$ & $9.86\times 10^{9}$ \\
fR-nu0.4 & $\bar{f}_{R0} = -1\times 10^{-4}$ & $0.4$ & $0.308$ & $0.0095$ & $6.4\times 10^{11}$ & $1.97\times 10^{10}$  \\
fR-nu0.6 & $\bar{f}_{R0} = -1\times 10^{-4}$ & $0.6$ & ${ 0.3032}$ & $0.0143$ & $6.27\times 10^{11}$ & $2.96\times 10^{10}$ \\
\hline
\end{tabular}
\caption{The suite of cosmological N-body simulations considered in the present work, with their main physical and numerical parameters. { Here $\Omega _{\rm CDM}$ and $\Omega _{\nu }$ indicate the energy density ratio to the critical density of the universe for CDM and neutrinos, respectively, while $M^{p}_{\rm CDM}$ and $M^{p}_{\nu }$ stand for the mass of CDM and neutrino particles in the different N-body runs.}}
\label{tab:models}
\end{table*}

\section{$f(R)$ modified gravity}
\label{foR}

In the present work, we will consider Modified Gravity cosmologies in the form
of $f(R)$ extensions of Einstein's General Relativity ({hereafter} GR). Such models are characterised by 
the action
\begin{equation}
\label{fRaction}
  S = \int {\rm d}^4x \, \sqrt{-g} \left( \frac{R+f(R)}{16 \pi G} + {\cal L}_m \right),
\end{equation}
where the Ricci scalar $R$ that appears in the standard Einstein-Hilbert action of GR has been
replaced by $R+f(R)$. In Eq.~(\ref{fRaction}), $G$ is Newton's gravitational constant, ${\cal L}_m$ is the
Lagrangian density of matter, and $g$ is the determinant of the metric
tensor $g_{\mu \nu }$. 
{ These models have been widely investigated in the literature, both in terms of their theoretical predictions at linear \citep[see e.g.][]{Pogosian_Silvestri_2008} and nonlinear \citep[see e.g.][]{Oyaizu_etal_2008,Schmidt_etal_2009,Li_etal_2012,Puchwein_Baldi_Springel_2013,Llinares_Mota_Winther_2013} scales, and in terms of
possible present \citep[see e.g.][]{Lombriser_etal_2012} and forecasted future \citep[][]{Zhao_etal_2009b} observational constraints.}
Within such framework, the quantity $f_R \equiv {\rm d}f(R)/{\rm d}R$ represents an additional
scalar degree of freedom obeying an independent dynamic equation that in the quasi-static approximation
and in the limit $f_{R} \ll 1$ takes the form\footnote{Whenever not explicitly stated otherwise, we always work in units where the speed of light is assumed to be unity, $c=1$.} { \citep[see again][]{Hu_Sawicki_2007}:}
\begin{equation}
  \nabla^2 f_R = \frac{1}{3}\left(\delta R - 8 \pi G \delta \rho \right) \,,
\label{eq:fR_field_eq}
\end{equation}
where $\delta R$ and $\delta \rho$ are the perturbations in the scalar curvature and
matter density, respectively.

Within the range of possible functional forms of $f(R)$, we will focus on the specific realisation proposed by \citet{Hu_Sawicki_2007}
which has the advantage of providing a close match to the standard $\Lambda $CDM background expansion history.
In { this} model, the function $f(R)$ takes the form
\begin{equation}
\label{fRHS}
f(R) = -m^2 \frac{c_1 \left(\frac{R}{m^2}\right)^n}{c_2 \left(\frac{R}{m^2}\right)^n + 1},
\end{equation}
where $ m^2 \equiv H_0^2 \Omega _{\rm M}$, with $H_0$ being the present-day
Hubble parameter and $\Omega _{\rm M}$ the dimensionless matter density parameter defined as
the ratio between the mean matter density and the critical density of the universe. In Eq.~(\ref{fRHS})
$c_{1}$, $c_{2}$, { and $n$} are non-negative constant free parameters that fully specify the model.
%$\Omega _{M}\equiv 8\pi G \rho _{M}/3H_{0}^{2}$.
By imposing the condition $c_2 (R/m^2)^n \gg 1$,
Eq.~(\ref{fRHS}) becomes $f(R) = -m^2
c_1 / c_2 + O((m^2/R)^n)$. 
It is then possible to obtain a background evolution close to that of a $\Lambda $CDM model
by setting the term $-m^2 c_1 /
c_2$ equal to $-2\Lambda$, where $\Lambda$ is the desired cosmological
constant, which is equivalent to { imposing} a relation between the two free parameters $c_{1}$ and $c_{2}$:
\begin{equation}
  \frac{c_1}{c_2} = 6 \frac{\Omega_\Lambda}{\Omega_{\rm M}},
\label{eq:c1c2_lambda}
\end{equation} 
where $\Omega_\Lambda \equiv \Lambda /3H_{0}^{2}$ is the present dimensionless density parameter of the cosmological constant.
Under { these} assumptions, the scalar degree of freedom $f_{R}$ takes the approximate form:
\begin{equation}
  f_R \approx -n \frac{c_1}{c_2^2}\left(\frac{m^2}{R}\right)^{n+1}.
\label{eq:fR-R,n_relation}
\end{equation}
By evaluating the present average scalar curvature of the universe $\bar{R}_{0}$ within a standard $\Lambda $CDM cosmology
and combining with Eq.~(\ref{eq:fR-R,n_relation}) one gets an equation for the background value of the present scalar
degree of freedom $\bar{f}_{R0}$. {Then, by} fixing the value of $\bar{f}_{R0}$ it is possible to obtain an independent relation
between $c_{1}$ and $c_{2}$ { as a function of $n$}, that combined with Eq.~(\ref{eq:c1c2_lambda}) completely determines both { parameters}. Therefore, the model
can be fully specified by fixing $n$ and $\bar{f}_{R0}$ rather than $c_{1}$ and $c_{2}$. Following the convention adopted in 
most of the literature we will stick to the former choice throughout the rest of the paper.

The above equations further simplify after fixing the value of $n$, which is set to unity in all the cosmological realisations presented in this work.
For $n=1$, then, the model is fully specified by the single parameter $\bar{f}_{R0}$, and the time evolution of the background scalar degree of freedom $\bar{f}_{R}$ is given by:
\begin{equation}
\bar{f}_{R}(a) = \bar{f}_{R0}
\left(\frac{\bar{R}_0}{\bar{R}(a)}\right)^2 = \bar{f}_{R0}
\left(\frac{1 + 4 \frac{\Omega_\Lambda}{\Omega_{\rm M}}}{ a^{-3} + 4
\frac{\Omega_\Lambda}{\Omega_{\rm M}}}\right)^2, 
\label{eq:fR_mean}
\end{equation}
while the curvature perturbation takes the form
\begin{equation}
\delta R = \bar{R}(a)\left(\sqrt{\frac{\bar{f}_{R}(a)}{f_{R}}} - 1\right). 
\label{eq:dR_fR}
\end{equation}

In $f(R)$ models, the gravitational potential $\Phi $
satisfies \citep{Hu_Sawicki_2007}
\begin{equation}
  \nabla^2 \Phi = \frac{16 \pi G}{3} \delta \rho - \frac{1}{6}\delta R \,,
\label{eq:phi_poisson_eq}
\end{equation}
where the second term on the right-hand side is responsible for the spatial dependence of the fifth-force arising as a consequence of the modification 
of the laws of gravity in addition to the standard Newtonian force. To carry out our analysis, we will resort { to} the
{\small MG-GADGET} implementation \citep[][]{Puchwein_Baldi_Springel_2013} of $f(R)$ models into the widely-used
parallel Tree-PM N-body code {\small GADGET3} that will be briefly described in Section~\ref{sims}. { This} numerical
tool allows to solve Eq.~(\ref{eq:fR_field_eq}) for a generic density distribution given by a set of N-body particles, and then compute the
total force arising on each particle through Eq.~(\ref{eq:phi_poisson_eq}), by including in the source term the curvature perturbation $\delta R$ 
obtained according to Eq.~(\ref{eq:dR_fR}).
We refer the interested reader to the {\small MG-GADGET} code paper for a more detailed presentation of the numerical implementation.

\begin{table}
\begin{center}
\begin{tabular}{cc}
\hline
Parameter & Value\\
\hline
$H_{0}$ & 67.1 km s$^{-1}$ Mpc$^{-1}$\\
$\Omega _{\rm M} $ & 0.3175 \\
$\Omega _{\rm DE} $ & 0.6825 \\
$ \Omega _{b} $ &0.049 \\
\hline
${\cal A}_{s}$ & $2.215 \times 10^{-9}$\\
$n_{s}$ & 0.966\\
\hline
\end{tabular}
\end{center}
\caption{The set of cosmological parameters adopted in the present work, consistent with the latest results of the Planck collaboration \citep[][]{Planck_016}. { Here $n_{s}$ is the spectral index of primordial density perturbations while ${\cal A}_{s}$ is the amplitude of scalar perturbations at the redshift of the CMB.}}
\label{tab:parameters}
\end{table}

\section{Massive Neutrinos and structure formation}
\label{intro_neutrinos}

Neutrinos were initially postulated in 1930 by Pauli as a solution to the apparent violation of energy,
momentum and spin { conservation} in the $\beta$-decay process. In 1956 their existence was
corroborated { for the first time} in the famous experiment carried out by Cowan \& Reines \citep[][]{Cowan_etal_1956}.
The measurements of the Z boson lifetime have { then} pointed out that the number of light neutrino 
families is three. Moreover, since the discovery of the neutrino 
oscillation phenomena \citep[][]{Cleveland_etal_1998} it is known that at least two of the
three neutrino families are massive, in contrast to the particle standard model assumption. 
Whereas measuring the absolute masses of the neutrinos is very difficult, the 
neutrino oscillations can be used to measure the mass square differences among the neutrino 
mass eigenstates. The most recent measurements from solar, atmospheric and reactor 
neutrinos result in: $\bigtriangleup m^2_{12}=7.5\times10^{-5}$ ${\rm eV}^2$ and $|\bigtriangleup 
m^2_{23}|=2.3\times10^{-3}$  ${\rm eV}^2$ \citep[][ respectively]
{Fogli_etal_2012,Forero_Tortola_Valle_2012}, with $m_1$, $m_2$ 
and $m_3$ being the masses of the three neutrino mass eigenstates. Unfortunately, the experimental
measurements do not allow us to { determine} the sign of the quantity $\bigtriangleup m^2_{23}$.
This gives rise to two different { possible} mass orderings (or hierarchies): the {\em normal} hierarchy where
$m_1< m_2 < m_3$, and the {\em inverted} hierarchy where $m_3< m_1 < m_2$. 

Knowing the absolute
masses of the neutrinos is of { critical} importance, since they represent a door to look for physics 
beyond the standard model. Whereas the neutrino oscillations have been used to set lower
limits on the sum of the neutrino masses, $\Sigma_i m_{\nu _i}\geq $  0.056 and 0.095 eV for the normal
and inverted hierarchies respectively, the tighter upper bounds on the masses of the neutrinos come
from cosmological observations.

In the very early universe neutrinos were in thermal equilibrium with the photons, the electrons and then positrons. While in thermal equilibrium, their momentum distribution followed
the standard Fermi-Dirac distribution. Then, it is easy to show that once neutrinos decouple from the primordial
plasma their momentum distribution follows 
\begin{equation}
n(p){\rm d}p=\frac{4\pi g_\nu}{(2\pi\hbar c)^3}\frac{p^2{\rm d}p}{e^\frac{p}{k_BT_\nu}+1}~,
\label{eq1}
\end{equation}
where $n(p)$ is the number of cosmic neutrinos with momentum between $p$ and $p+{\rm d}p$, 
$g_\nu$ is the number of neutrino spin states { and  $k_B$ is the Boltzmann constant. 
The temperature of the cosmic neutrino background and that of the
CMB are related by $T_\nu(z=0)=\left(\frac{4}{11}\right)^{1/3}T_\gamma(z=0)$ 
 \citep[see for instance][]{Weinberg_book} such that the temperature of the neutrino cosmic background at redshift
$z$ is given by $T_\nu(z)\cong1.95(1+z)$ K.} { The current abundance of relic neutrinos, obtained by integrating Eq.~(\ref{eq1}), results in 113 neutrinos per cubic centimeter per neutrino family.} The fraction of the total 
energy density of the universe made up by massive neutrinos can also be easily derived 
from Eq.~(\ref{eq1}):
\begin{equation}
\label{omega_nu_eq}
\Omega_\nu=\frac{\Sigma_i m_{\nu_i}}{93.14h^2 {\rm eV}}~.
\end{equation}

During the radiation-dominated era, neutrinos constituted an important fraction of the total energy 
density of the universe, playing a crucial role in setting the abundance of the primordial elements. 
The impact of massive neutrinos on cosmology is very well understood 
at the linear order \citep[][]{Lesgourgues_Pastor_2006}: 
on one hand, massive neutrinos modify the matter-radiation equality time, and on the other hand, 
they slow down the growth of matter perturbations. The combination of the above two effects produces 
a suppression on the matter power spectrum { at}  small scales \citep[see for instance][]
{Lesgourgues_Pastor_2006}. The signatures left by neutrino masses on both the matter 
power spectrum and the CMB have been widely used to set upper limits { on} their masses.
Different studies point out that $\Sigma_i m_{\nu_i} < 0.3$ eV with a confidence level of 
$2\sigma$ \citep{Zhao_etal_2013, Xia_etal_2012, RiemerSorensen_etal_2012,
Joudaki_2012,Planck_016}.

Even though the impact of neutrino masses is well understood at the linear order, 
the impact of massive neutrinos on the fully non-linear regime { has so far received little attention}.
Recent studies using N-body simulations with massive neutrinos have investigated the 
neutrino imprints on the nonlinear matter power spectrum \citep[][]{Brandbyge_etal_2008, Brandbyge_Hannestad_2009, Brandbyge_Hannestad_2010, 
Viel_Haehnelt_Springel_2010, Bird_Viel_Haehnelt_2012, Agarwal_Feldman_2011, 
Wagner_Verde_Jimenez_2012}. Such studies have pointed out that
in the fully non-linear regime massive neutrinos induce a scale- and redshift-dependent 
suppression on the matter power spectrum which can be up to $\sim25\%$ 
larger than the predictions of the linear theory. 

The mean thermal velocity of the cosmic neutrinos, obtained from Eq.~(\ref{eq1}), is given by 
$\sim160(1+z)\frac{{\rm eV}}{m_\nu}$ km/s. For neutrinos with $\Sigma_i m_{\nu_i}$ = 0.3 eV
their mean velocity today is then $\sim500$ km/s, i.e. low enough to cluster within the gravitational
potential wells of galaxy clusters. The neutrino clustering has been studied in several works
\citep[][]{Singh_Ma_2003, Ringwald_Wong_2004, Brandbyge_etal_2010, 
Villaescusa-Navarro_etal_2011, Villaescusa-Navarro_etal_2013}. In 
\citet{Villaescusa-Navarro_etal_2011}, for instance, 
the authors studied the possibility of detecting the excess of mass
in the outskirts of galaxy clusters due to the extra mass contribution arising from the clustering
of massive neutrinos. Furthermore, in 
\citet{Villaescusa-Navarro_etal_2013} it was shown that the neutrino clustering also affects the
shape and amplitude of the neutrino momentum distribution of Eq.~(\ref{eq1}) at low redshift.

Massive neutrinos also leave their imprint on the halo mass function (HMF). In the pioneering work of
\citet{Brandbyge_etal_2010} the authors computed the HMF in cosmologies with massive neutrinos
using N-body simulations. They compared their results with the HMF obtained by using the 
Sheth-Tormen HMF { and} found that reasonable agreement between the results of the N-body
simulations and the Sheth-Tormen HMF \citep[][]{Sheth_Tormen_1999} was achieved if the latter was computed using the total
matter power spectrum but the mean CDM density of the universe, $\rho_{\rm CDM}$. { These} findings
were subsequently corroborated by \citet{Villaescusa-Navarro_etal_2013, Marulli_etal_2011}. However, 
in the recent work by \citet{Ichiki_Takada_2012} it was suggested that an even better agreement 
would be obtained by using the linear CDM power spectrum instead of the total matter linear power
spectrum. This claim has been tested against N-body simulations resulting in an excellent agreement 
\citep[see e.g.][]{Castorina_etal_2013, Costanzi_etal_2013}. Moreover, in \citet{Castorina_etal_2013} it was shown that this is the only way of parameterising the HMF in an almost universal form, i.e. independently of the underlying cosmological model.

The impact of neutrino masses on the {LSS} of the universe covers a wide range of 
aspects. In \citet{Villaescusa-Navarro_etal_2012} it was suggested that massive neutrinos 
could play an important role in the dynamics of cosmological voids. The reason is that the large
thermal velocities of the neutrinos make them less sensitive to the gravitational evolution of voids. 
In other words, one would naively expect cosmic voids to be empty of CDM, but not of cosmological
neutrinos. The contribution of neutrinos to the void's total mass has then important consequences
in terms of its evolution. In \citet{Villaescusa-Navarro_etal_2012} the authors used such argument
to study the possible signature left by massive neutrinos on Lyman-$\alpha$ voids.

In a series of recent papers \citep[][]{Villaescusa-Navarro_etal_2013b, Castorina_etal_2013} it has also been
shown that on large scales the bias between the spatial distribution of dark matter halos and that of the 
underlying matter exhibits a dependence on scale for cosmological models with massive neutrinos. 
Such scale-dependence is highly suppressed if the bias is computed with respect to the spatial 
distribution of the CDM instead of the total matter.\\

As briefly summarised in the present section, the effects induced by massive neutrinos on the {LSS} of the universe are 
numerous. In this work we further explore such effects in non-standard cosmologies. More specifically, we study the combined effects on structure formation of
a cosmic background of massive neutrinos with different values of the total neutrino mass and of a modification of standard gravity in the form of the $f(R)$ gravitational action introduced in Eq.~(\ref{fRaction}). The main aim of the present work is therefore to investigate whether the simultaneous existence of an extended theory of gravity and of a 
{ massive} neutrino background might impact on the observational capability of detecting and/or constraining any of such two extensions of the standard cosmological scenario. Then, as will become clear in the rest of the paper, the upper bounds on the total neutrino mass { quoted in the present section} that have been obtained through cosmological observations might not be valid anymore if the underlying theory of gravity is different from standard GR. Analogously, present and future constraints on the nature of gravity might be strongly biased in the presence of a significant massive neutrino background.

\begin{figure}
\includegraphics[scale=0.45]{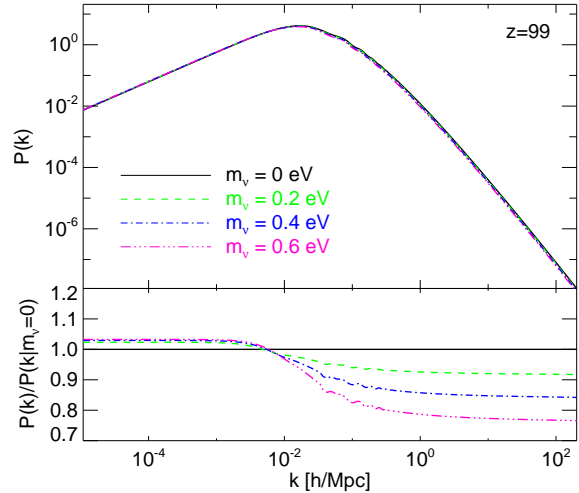}
\caption{{ The linear matter power spectra adopted for the initial conditions of the simulations with different total neutrino mass.}}
\label{fig:linear_power}
\end{figure}

\section{Cosmological simulations}
\label{sims}

The aim of the present work is to investigate the joint effects of Modified Gravity and massive neutrinos on the basic statistical properties of structure formation, like the nonlinear matter power spectrum, the halo mass function, and the large-scale halo bias, with the goal of quantifying the level of degeneracy between these two independent extensions of the standard cosmological scenario that is usually adopted for large N-body simulations, i.e the $\Lambda $CDM cosmology with massless neutrinos. 
Furthermore, we aim to assess the discriminating power of present and future observational data with respect to such extended cosmologies. In order to include in our comparison both the linear and nonlinear regimes of structure formation to investigate whether a suitable combination of such different regimes might help in breaking -- or at least alleviating -- possible degeneracies, we need to rely on dedicated N-body simulations that include at the same time the effects of $f(R)$ modifications of gravity
and of a cosmic neutrino background with different possible values of the total neutrino mass $\Sigma _{i}m_{\nu _{i}}$. To this end, we have combined the {\small MG-GADGET} implementation of $f(R)$ models by \citet{Puchwein_Baldi_Springel_2013} with the massive neutrinos module developed by \citet{Viel_Haehnelt_Springel_2010}, both included in the widely-used Tree-PM N-body code {\small GADGET3}. 
{ More specifically, the simulations presented in this work are based on explicitly modelling the neutrinos with particles, as opposed to adopting the approximate Fourier-based technique (where they are added as an extra-force in k-space). Using particles has the disadvantage of having significant Poisson noise in the neutrino density field, but it is ultimately the more general technique capable of accounting correctly for all non-linear effects. Furthermore, our numerical setting ensures that noise issues are unimportant in the regime that we are analysing.}
{ The} combination { of these two different modules} has required { some} changes to the original independent implementations in order to allow the handling of very high memory allocation requirements arising when applying the multi-grid acceleration scheme of {\small MG-GADGET} \citep[see again][for a detailed description of the multi-grid scheme]{Puchwein_Baldi_Springel_2013} to two particle species with very different spatial distributions. 

\begin{figure}
\centering
\includegraphics[scale=0.43]{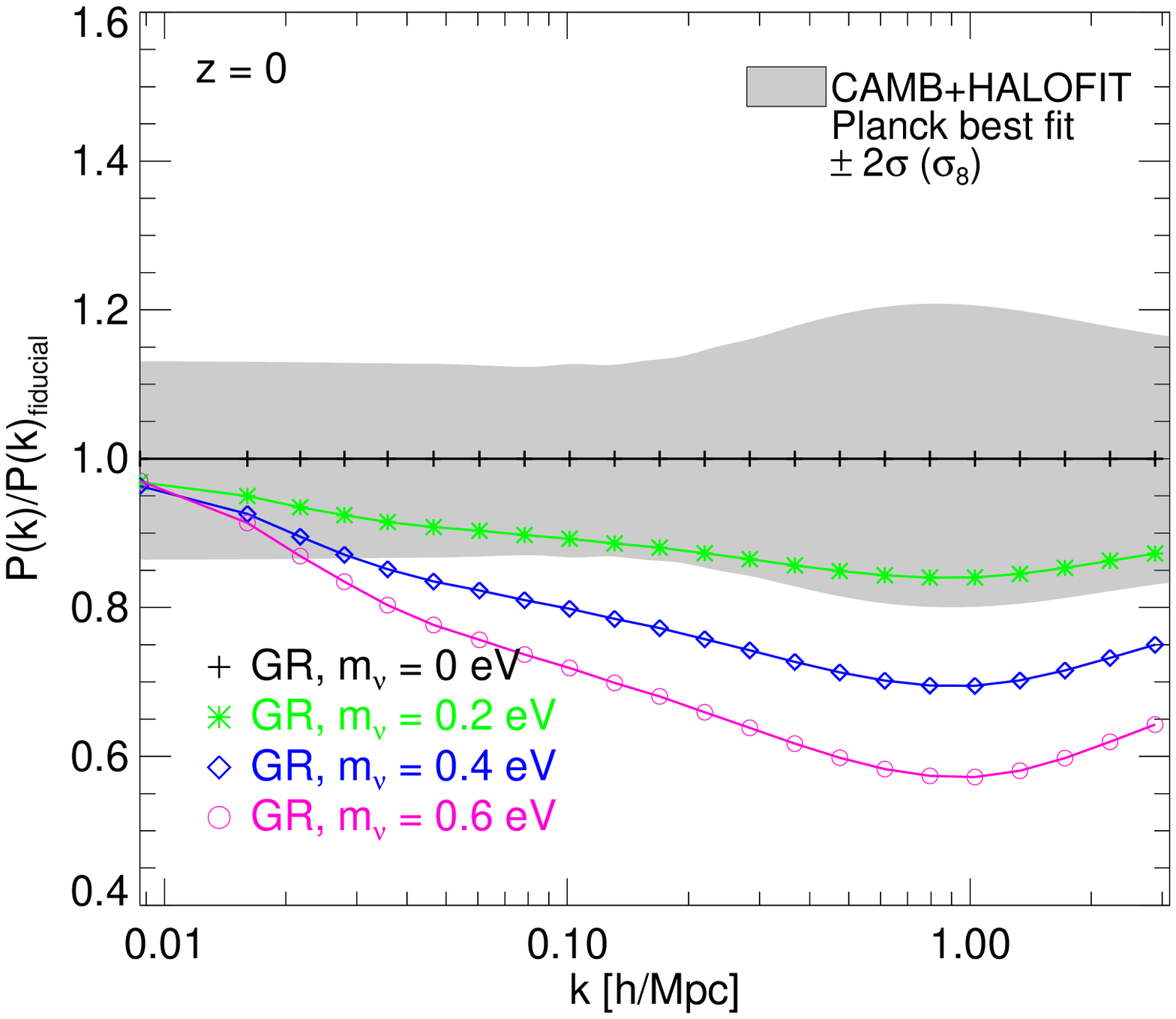}
\includegraphics[scale=0.43]{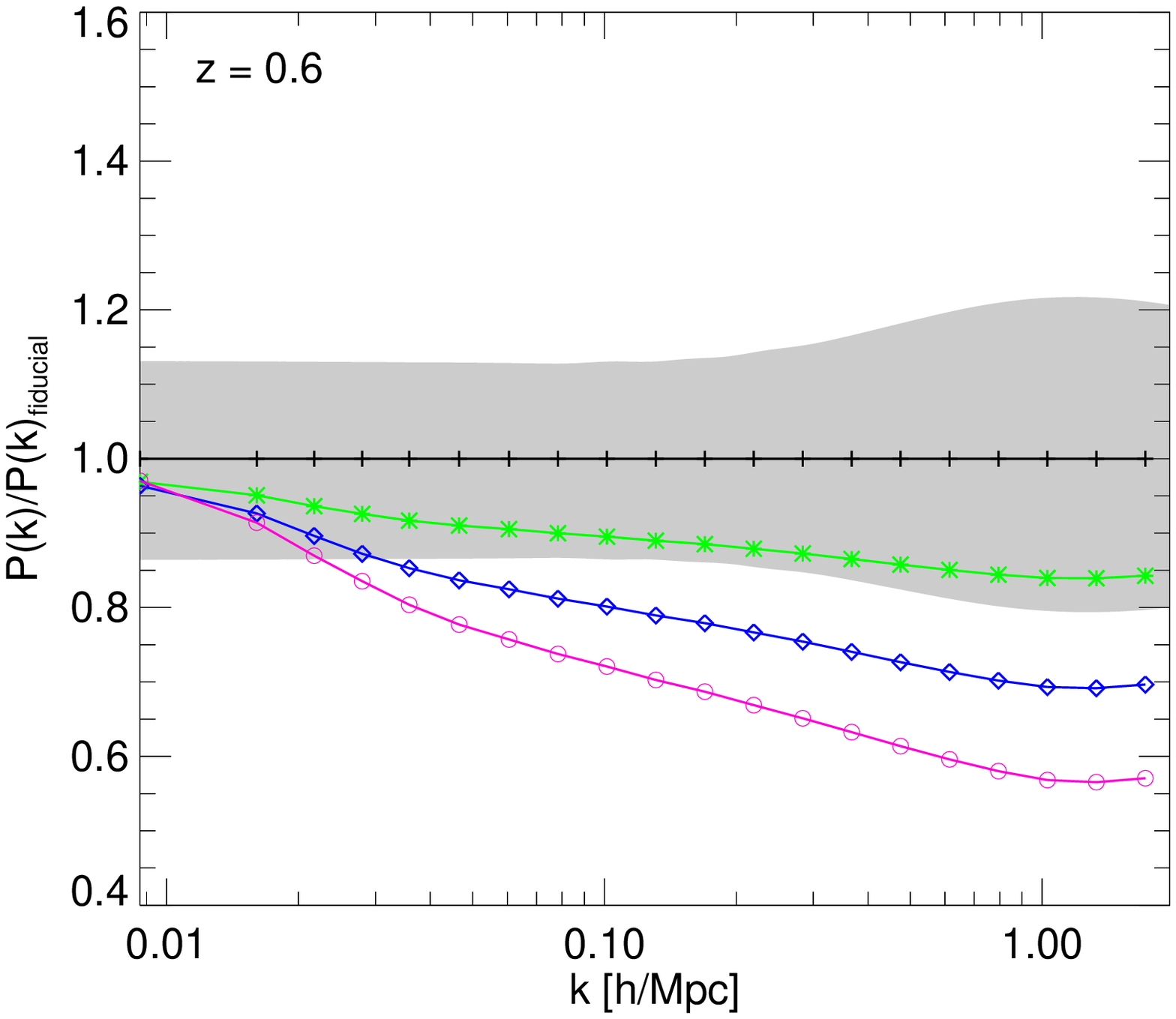}
\includegraphics[scale=0.43]{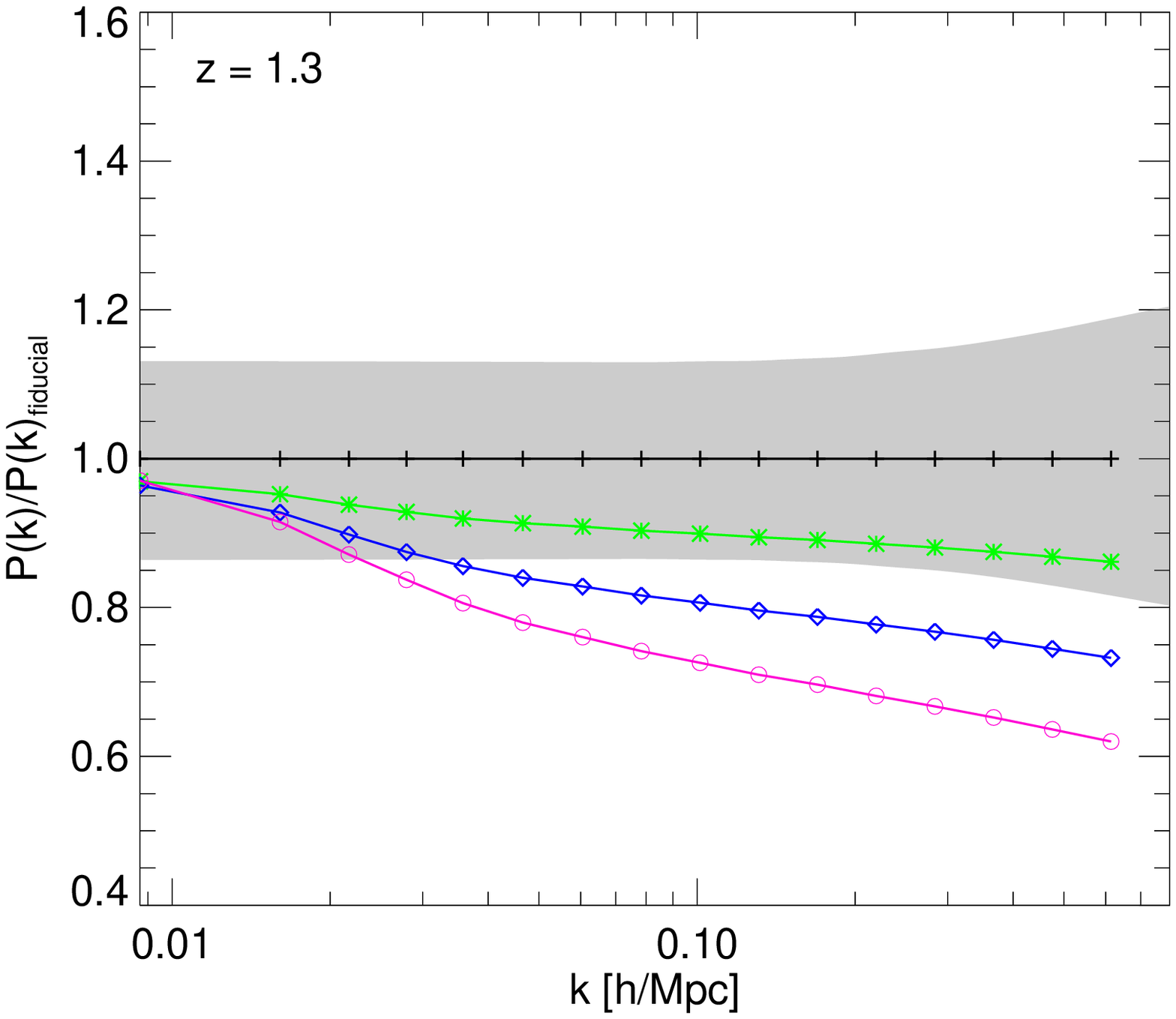}
\caption{The { nonlinear} matter power spectrum ratio with respect to the fiducial model for different values of the total neutrino mass { as labelled. The different panels refer to $z=0$ ({\em top}), $z=0.6$ ({\em middle}), and $z=1.3$ ({\em bottom}).} The grey shaded area represents the region obtained with {\small CAMB} and {\small HALOFIT} by setting all cosmological parameters to their fiducial Planck values except $\sigma _{8}$ which is allowed to vary within its 2-$\sigma $ confidence interval.}
\label{fig:power_GR}
\end{figure}
\begin{figure}
\centering
\includegraphics[scale=0.43]{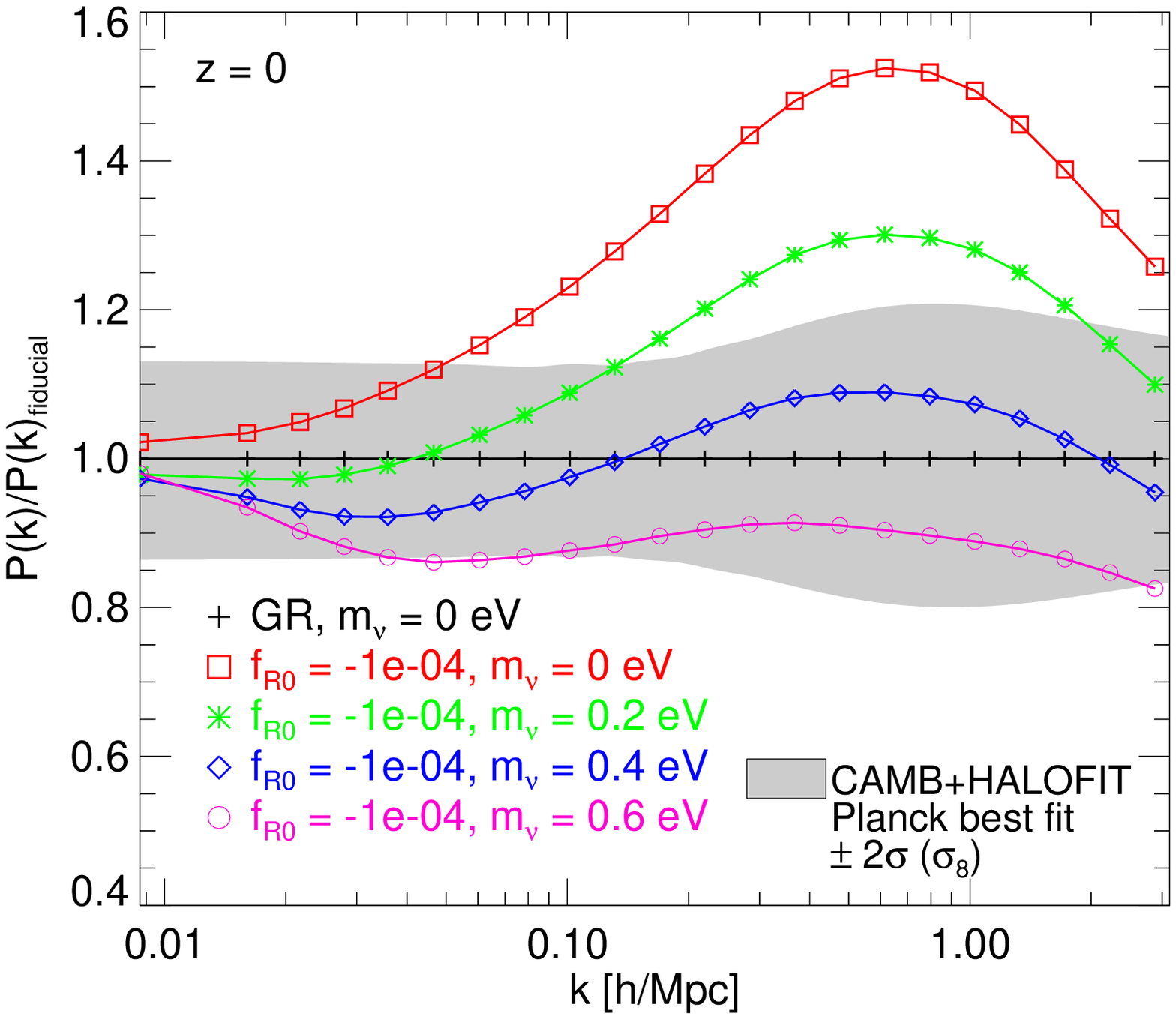}
\includegraphics[scale=0.43]{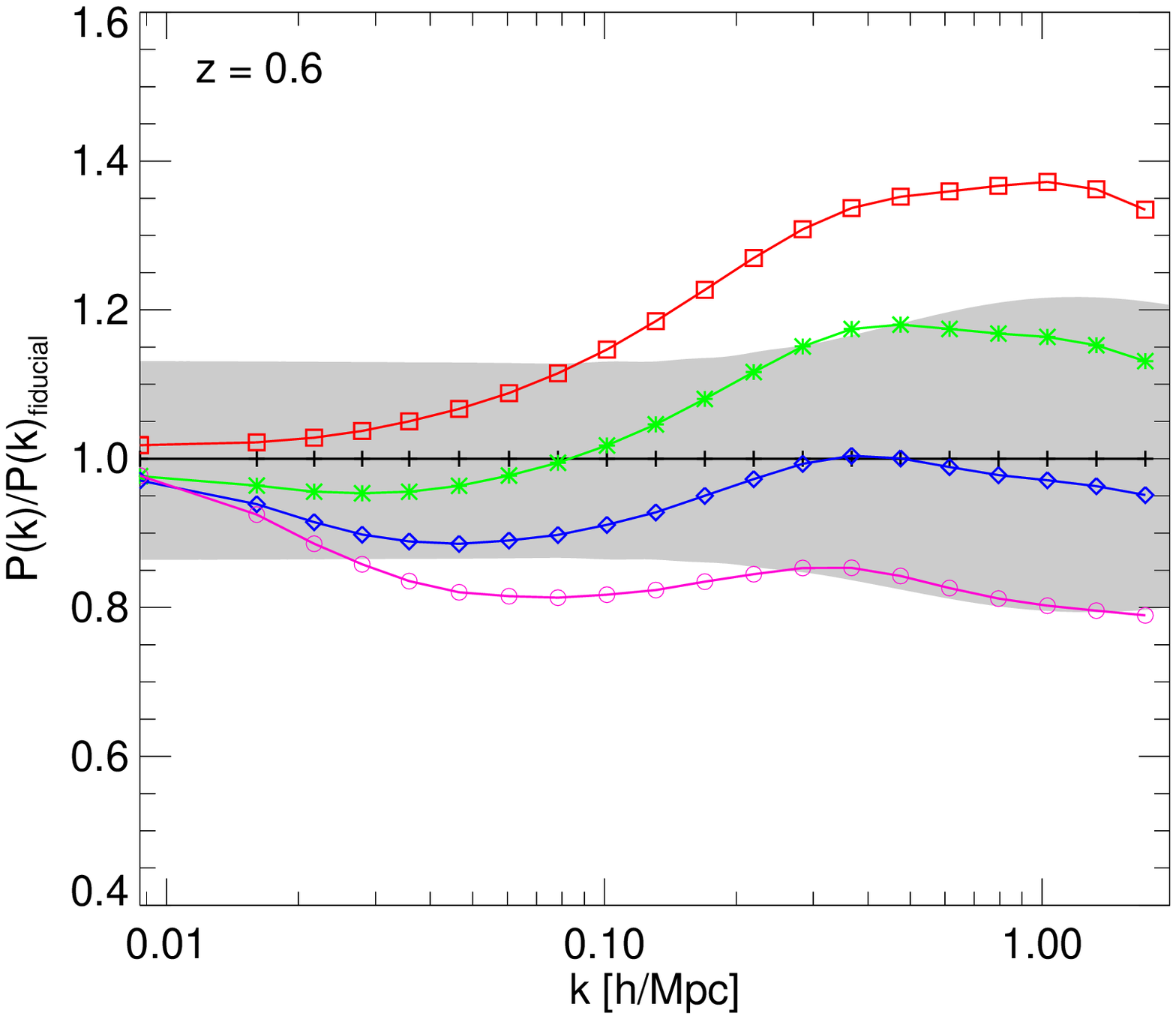}
\includegraphics[scale=0.43]{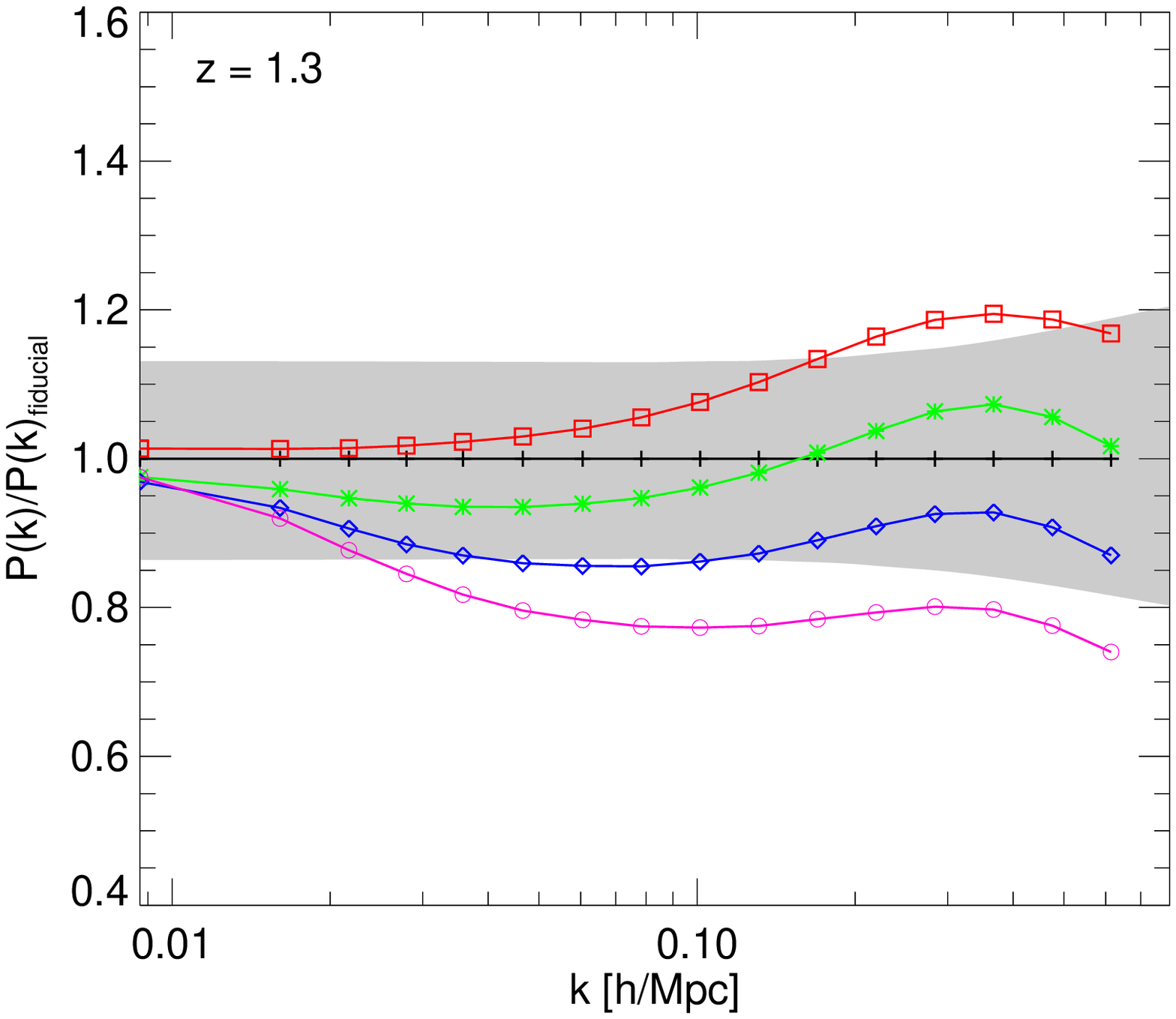}
\caption{{ As Fig.~\ref{fig:power_GR} but for the combined simulations of $f(R)$ gravity and massive neutrinos.}\\
\\
\\}
\label{fig:power_fR}
\end{figure}

\begin{figure*}
\includegraphics[scale=0.32]{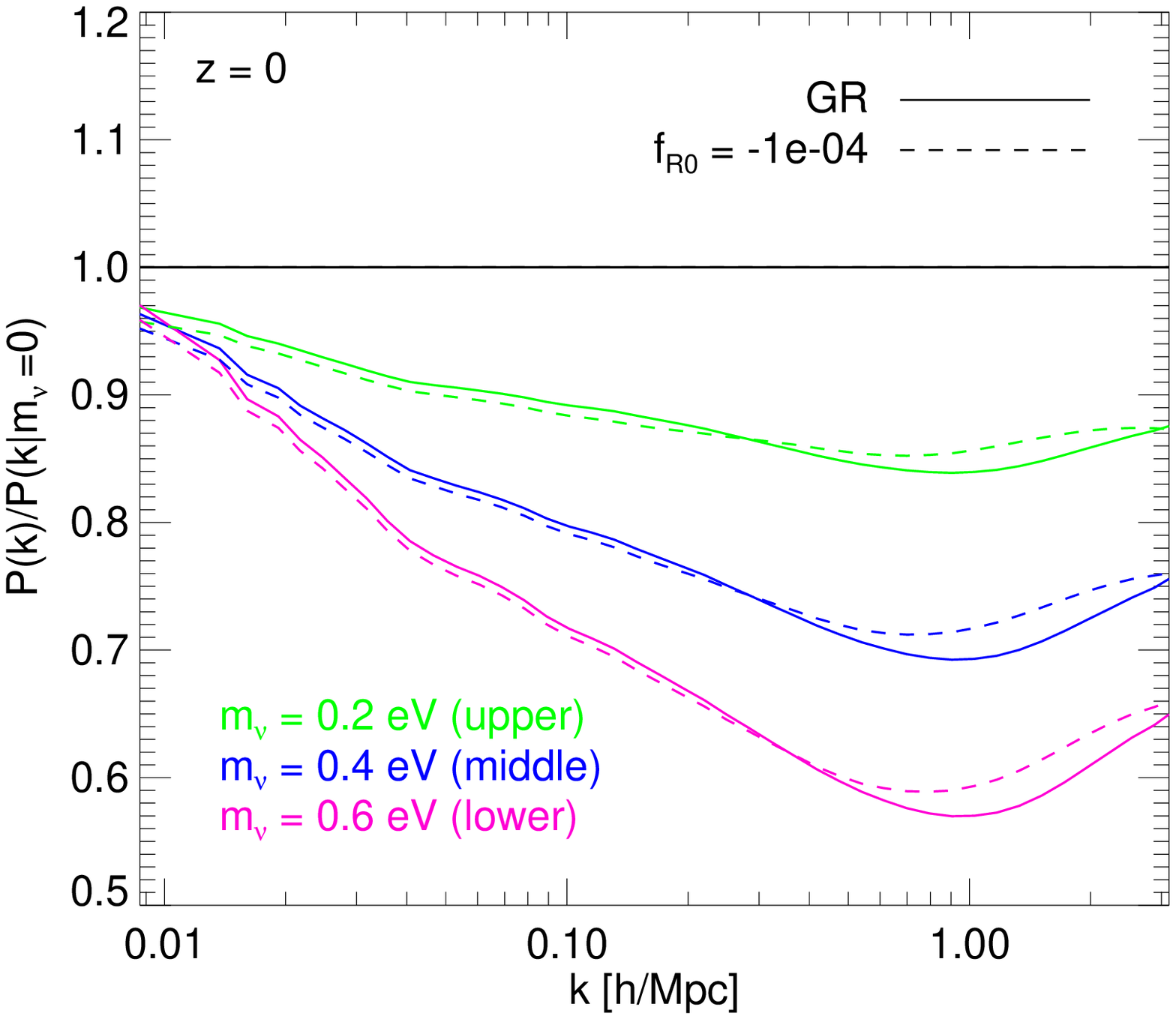}
\includegraphics[scale=0.32]{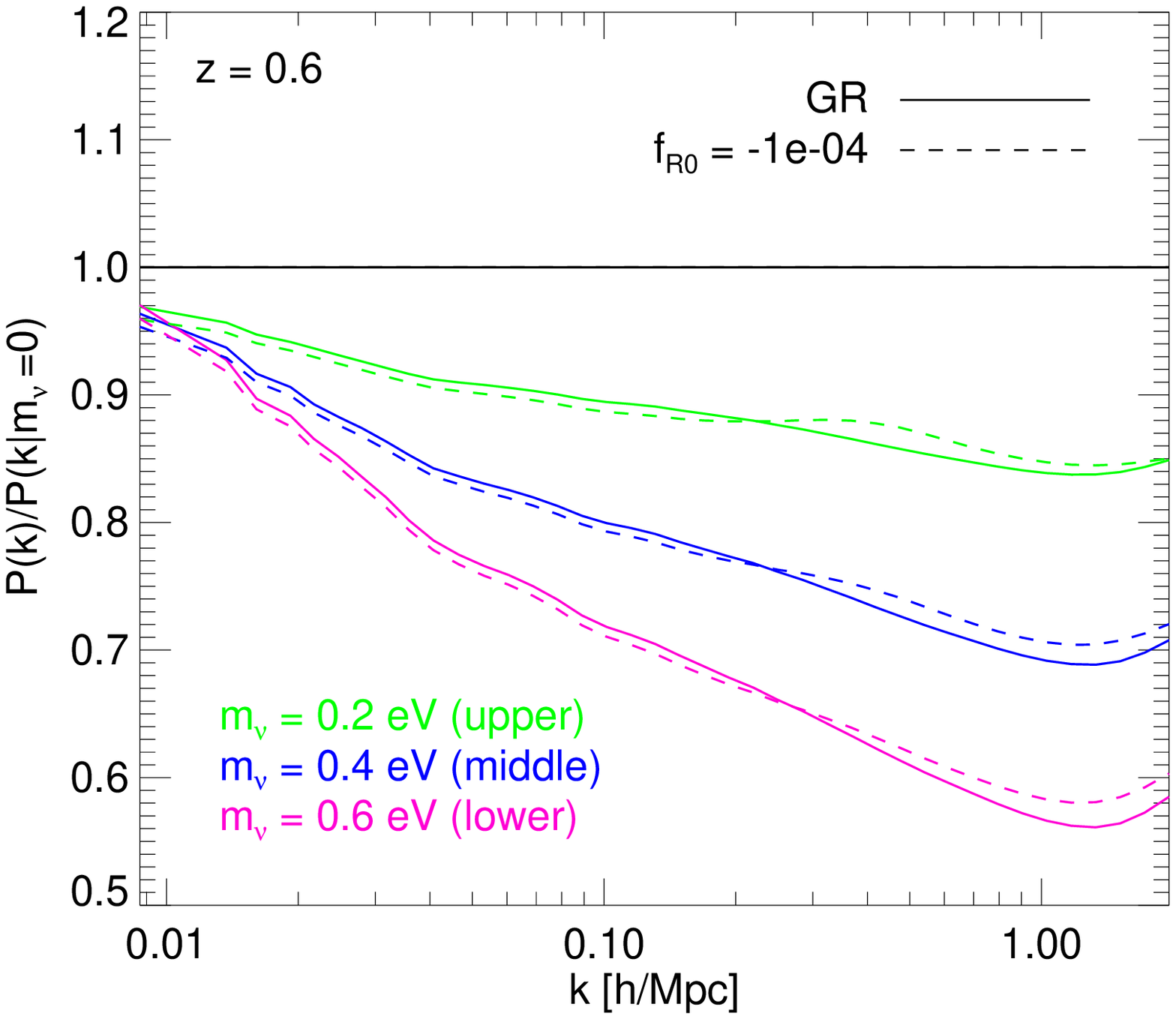}
\includegraphics[scale=0.32]{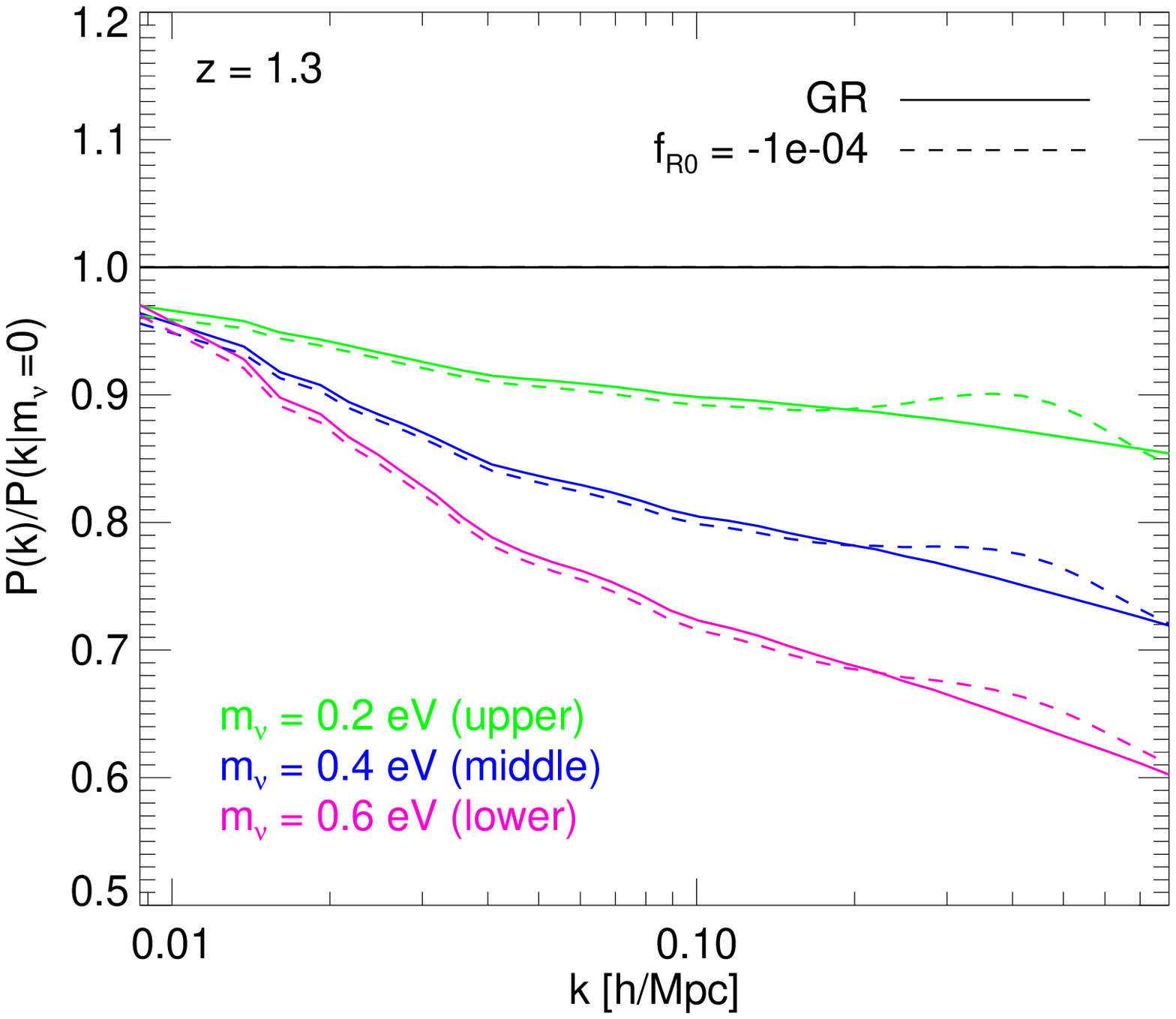}
\caption{The { matter} power spectrum suppression due to the presence of massive neutrinos in both GR (solid lines) { and} $f(R)$ (dashed lines) models as compared to the massless neutrino case at different redshifts{: $z=0$ ({\em left}), $z=0.6$ ({\em middle}), and $z=1.3$ ({\em right})}.}\label{fig:power_comparison}
\label{fig:power_comparison}
\end{figure*}

\begin{figure*}
\includegraphics[scale=0.32]{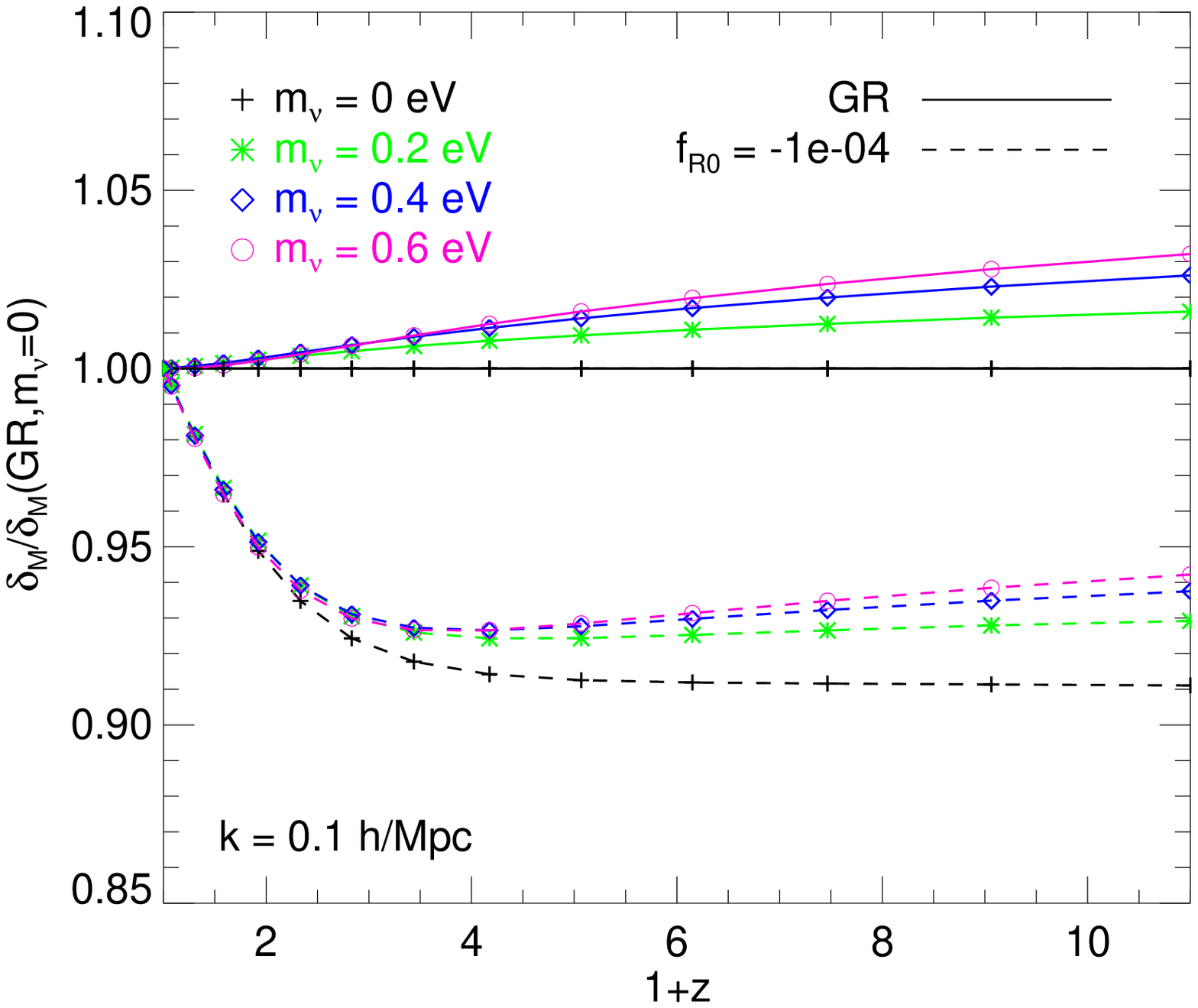}
\includegraphics[scale=0.32]{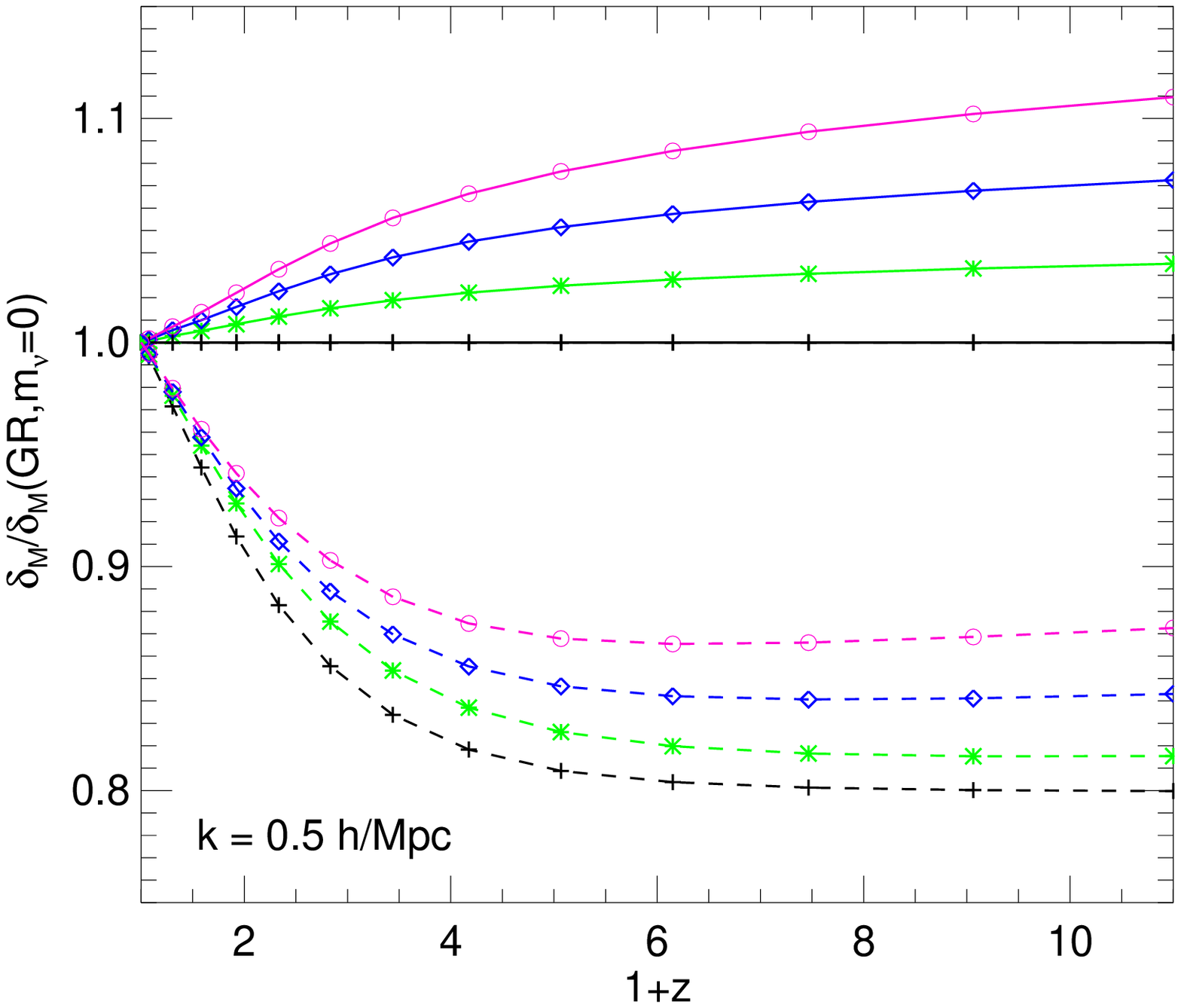}
\includegraphics[scale=0.32]{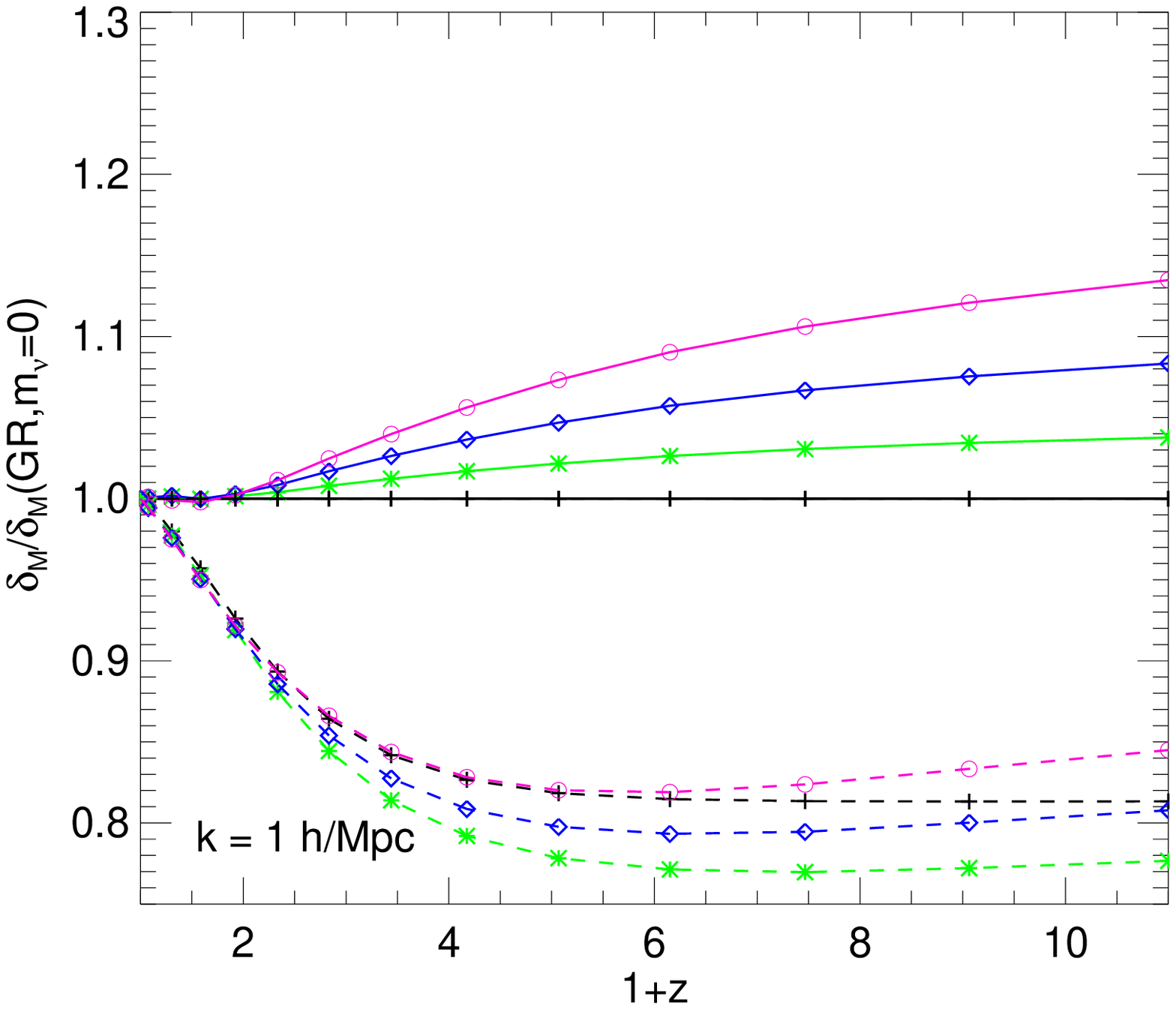}
\caption{The evolution of the matter density perturbations amplitude as a function of redshift normalised to the fiducial model at three different scales in the linear ($k=0.1\, h/$Mpc, {\em left}), mildly-nonlinear ($k=0.5\, h/$Mpc, {\em middle}), and nonlinear ($k=1\, h/$Mpc, {\em right}) regimes of structure formation.}
\label{fig:growth_rate}
\end{figure*}

With such combined N-body code at hand, we have performed a suite of intermediate-resolution N-body simulations on a periodic cosmological box of $1$ Gpc$/h$ aside filled with an equal number $N=512^3$ of CDM and neutrino particles, for a range of different cosmological models characterised by two possible theories of gravity -- namely standard GR and $f(R)$ with $n=1$ and $\bar{f}_{R0}=-1\times 10^{-4}$ -- and different values of the total neutrino masses $\Sigma _{i}m_{\nu _{i}} = \left\{ 0\,, 0.2\,, 0.4\,, 0.6\right\} $ eV. The full range of cosmologies covered by our simulations suite is summarised in Table~\ref{tab:models}. All runs are carried out using the latest Planck constraints \citep[][]{Planck_016} on { the background cosmological parameters, which are summarised in the upper part of Table~\ref{tab:parameters}, and on the amplitude ${\cal A}_{s}$ and spectral index $n_{s}$ of the linear matter power spectrum at $z_{\rm CMB}\approx 1100$ (listed in the bottom part of Table~\ref{tab:parameters}), thereby resulting in slightly different values of $\sigma _{8}(z=0)$ for the different cosmologies.}

Initial conditions have been generated with a modified version of the widely-used {\small N-GENIC} code that allows to include an additional matter 
component -- besides the standard CDM and gas particle types -- with the density perturbations and the thermal velocity distribution of a cosmic neutrino background as computed by the public Boltzmann code {\small CAMB}\footnote{www.cosmologist.info} \citep[][]{camb} for any given total neutrino mass $\Sigma _{i}m_{\nu _{i}}$. 
In particular, the initial conditions were generated at $z=99$ by perturbing the positions of the CDM and neutrino
particles, that were set initially into a cubic regular grid, using the Zel'dovich approximation. 
{ The linear matter power spectra employed to set up initial conditions for the various total neutrino masses are displayed in Fig.~\ref{fig:linear_power}, along with the ratio to the fiducial massless neutrino case. Since at such high redshift the MG fifth-force is fully screened, for each total neutrino mass we used the same initial conditions for both the GR and the $f(R)$ simulations.}

We incorporate the effects
of baryons by using a transfer function that is a weighted average of the CDM and baryon transfer
functions as given by CAMB, even if in the present simulations we do not include baryonic particles as a separate species in the N-body runs. 
The neutrino particles also receive an extra velocity component arising from 
random sampling the neutrino momentum distribution at the starting redshift. In all the different runs, the total matter density is kept fix { at} the
Planck value of $\Omega _{\rm M} = 0.3175$ which is split into the contribution of CDM and massive neutrinos, with the latter being given by Eq.~(\ref{omega_nu_eq}). 
As a result, the CDM density and the corresponding mass of simulated CDM particles decreases when $\Sigma _{i}m_{\nu _{i}}$ is increased, as summarised in Table~\ref{tab:models}.\\

\section{Results}
\label{res}

We now describe the main outcomes of our numerical investigation of combined $f(R)$ and massive neutrinos cosmologies, concerning the nonlinear matter power spectrum, the CDM halo mass function, and the halo{-matter} bias.

\subsection{The nonlinear matter power spectrum}

For each of our cosmological simulations, we have computed the total (i.e. CDM {} plus neutrinos) nonlinear matter power spectrum by determining the density field
on a cubic cartesian grid with twice the resolution of the PM grid used for the N-body integration (\ie $1024^{3}$ grid nodes) through a Cloud-in-Cell mass assignment. 
{ This} procedure provides a determination of the nonlinear matter power spectrum 
up to the Nyquist frequency of the PM grid, corresponding to $k_{\rm Ny} = \pi N/L \approx 3.2\, h/$Mpc.
The obtained power spectrum is then truncated at the $k$-mode where the shot noise reaches $20\%$ of the measured power. With { the} simulated power spectra, we can estimate the separate effects of $f(R)$ modified gravity and of massive neutrinos on the linear and nonlinear regimes of structure formation at different redshifts. Furthermore, we are able to compute for the first time the joint effects of both independent extensions of the standard fiducial cosmological model on the amplitude of linear and nonlinear density perturbations.

In Fig.~\ref{fig:power_GR} we display the ratio of the nonlinear matter power spectrum of standard GR cosmologies with different neutrino masses to the fiducial case of massless neutrinos, for { the} three different redshifts { $z=\left\{ 0\,, 0.6\,, 1.3\right\}$.} As expected, the {massive} neutrino component suppresses the power spectrum amplitude both at linear and nonlinear scales, with a progressively more significant impact for larger values of the { total} neutrino mass { $\Sigma _{i}m_{\nu _{i}}$}. This is consistent with numerous previous investigations of structure formation processes in the presence of a massive neutrino component \citep[see e.g.][]{Brandbyge_etal_2008,Viel_Haehnelt_Springel_2010,Bird_Viel_Haehnelt_2012,Agarwal_Feldman_2011,Wagner_Verde_Jimenez_2012}.

In the uppermost panel of Fig.~\ref{fig:power_fR}, instead, we show the same power ratio to the fiducial model for the combined simulations of $f(R)$ MG with different neutrino masses at $z=0$. As one can see in the figure, for the massless neutrino run (red line) the power spectrum ratio shows the typical shape already found by several authors \citep[see e.g.][]{Oyaizu_etal_2008,Ecosmog,Puchwein_Baldi_Springel_2013} for the same $\bar{f}_{R0}=-1\times 10^{-4}$ MG scenario investigated here, { which is} characterised by a growing enhancement at linear and mildly nonlinear scales that reaches a peak of about $50\%$ at $k\sim 0.6\, h/$Mpc (at $z=0$), followed by a decrease at progressively more nonlinear scales. The latter decrease is produced by the shift of the transition scale between the 1-halo and 2-halo term domination in the density power spectrum associated { with} the effectively larger value of $\sigma _{8}$ in the $f(R)$ cosmology arising at low redshifts as a consequence of the MG fifth-force \citep[see e.g.][]{Zhao_Li_Koyama_2011a}.
{ Therefore,} our pure $f(R)$ run is { fully} consistent with previous findings for the same cosmological model and with the results of \cite{Puchwein_Baldi_Springel_2013} obtained with the {\small MG-GADGET} code. { The same quantity is displayed in the remaining panels of Fig.~\ref{fig:power_fR} for the same redshifts shown in Fig.~\ref{fig:power_GR}.}

One of the main results of the present paper is then represented by the remaining curves of Fig.~\ref{fig:power_fR} that refer to the impact on the matter power spectrum of different neutrino masses { (green, blue, and magenta for $\Sigma _{i}m_{\nu _{i}}=0.2\,, 0.4\,, 0.6$ eV, respectively)} in the context of an underlying $f(R)$ cosmology. As one can see from the figures, the effect of $f(R)$ on the matter power spectrum is strongly suppressed by the presence of massive neutrinos such that the MG cosmology with neutrino masses of $\Sigma _{i}m_{\nu _{i}} = 0.4$ eV does not deviate more than $\approx 10\%$ (at $z=0$) from the fiducial scenario. 
In both Figs.~\ref{fig:power_GR} and \ref{fig:power_fR} the grey shaded areas correspond to the ratio of the $\Lambda $CDM nonlinear matter power spectrum computed with the {\small HALOFIT} package \citep[][]{Smith_etal_2003} within the {\small CAMB} code by setting all cosmological parameters to their fiducial Planck values except for $\sigma _{8}$ which is allowed to vary within its 2-$\sigma $ confidence interval \citep[according to][]{Planck_016}. In other words, the grey shaded areas provide a visual estimate of the discriminating power of presently available { cosmological} data. { This is clearly just a rough indicative estimate since we are varying only one parameter, but still provides a quantitative idea of how the differences among the various models compare with deviations allowed by present observational uncertainties on standard cosmological parameters.}

These results clearly show { that} $f(R)$ MG models are strongly degenerate with massive neutrinos, such that a combination of both extensions of the standard cosmological model might prevent or significantly { dim the prospects} to constrain any of them through a direct observational determination of the matter power spectrum at low redshifts, { unless} an independent measurement of the total neutrino mass from particle physics { experiments becomes available. This} conclusion extends the previous results of \citet{He_2013} { (based on CMB observations at high-$z$)} { and the recent analysis of \citet{Motohashi_etal_2013} (which combines both high-$z$ CMB data and low-$z$  linear power spectrum measurements)} to the { low-$z$} nonlinear regime of structure formation by employing for the first time specifically designed N-body simulations that allow to consistently evolve $f(R)$ MG cosmologies in the presence of massive neutrinos. { This} degeneracy therefore poses a further theoretical limitation { on} the constraints that can be inferred on cosmological parameters and extended cosmological models from accurate observational determinations of the matter power spectrum, besides the widely discussed degeneracy with the uncertain effects of baryonic physical processes at small scales, such as e.g. the feedback from Active Galactic Nuclei \citep[see e.g.][for an estimate of the biasing effects of AGN feedback mechanisms on the determination of standard cosmological parameters and on the peculiar signatures of $f(R)$ MG models, respectively]{Semboloni_etal_2011,Puchwein_Baldi_Springel_2013}.
Nonetheless, as one can see by comparing the different panels of Fig.~\ref{fig:power_fR}, the redshift evolution of the power spectrum might provide a way to disentangle
the two effects if sufficiently precise observational measurements of the matter power spectrum up to a comoving wavelength of $k\approx 10\, h/$Mpc can be obtained for different values of {$z\lesssim 1.3$}.

\begin{figure}
\centering
\includegraphics[scale=0.45]{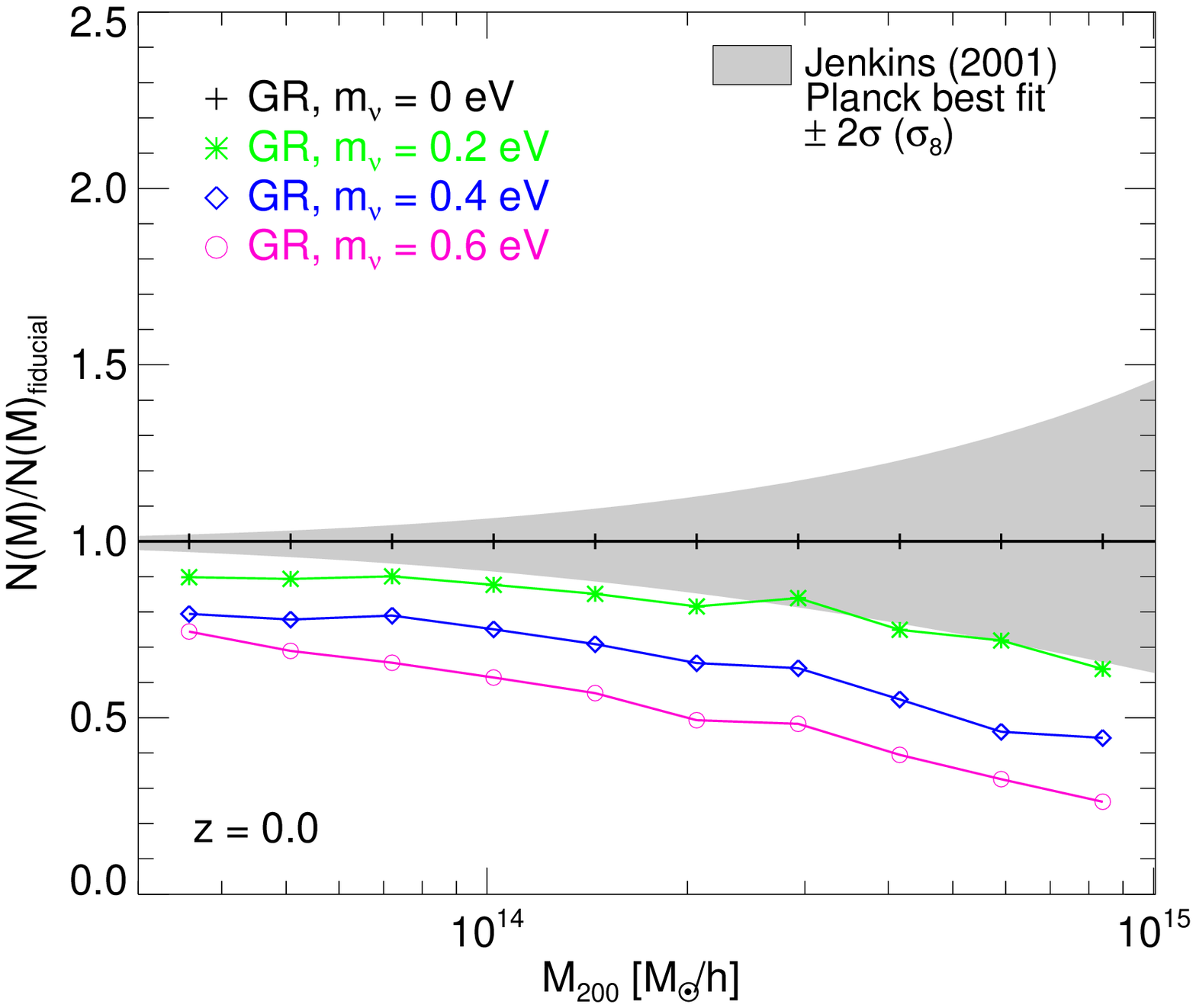}
\includegraphics[scale=0.45]{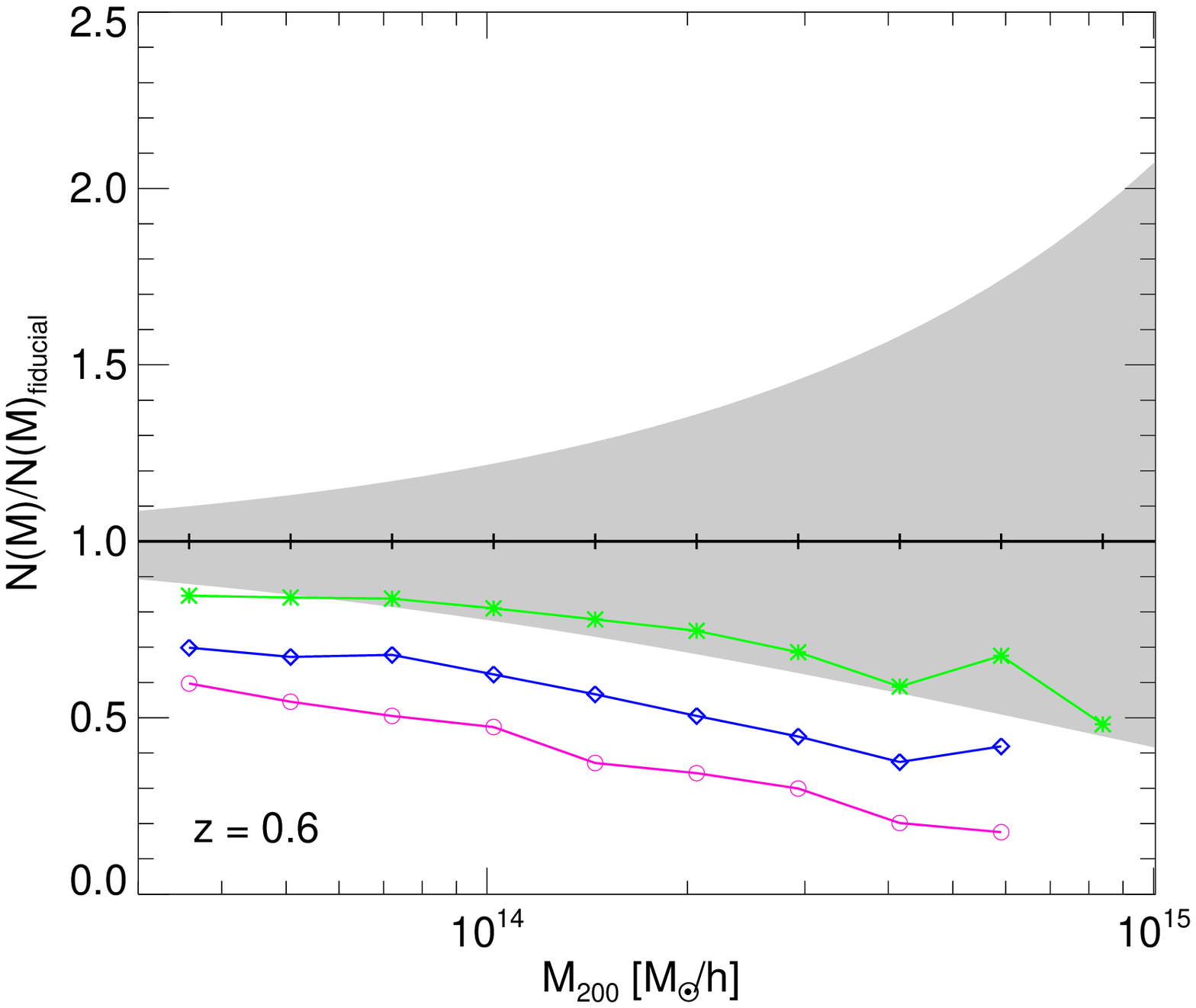}
\includegraphics[scale=0.45]{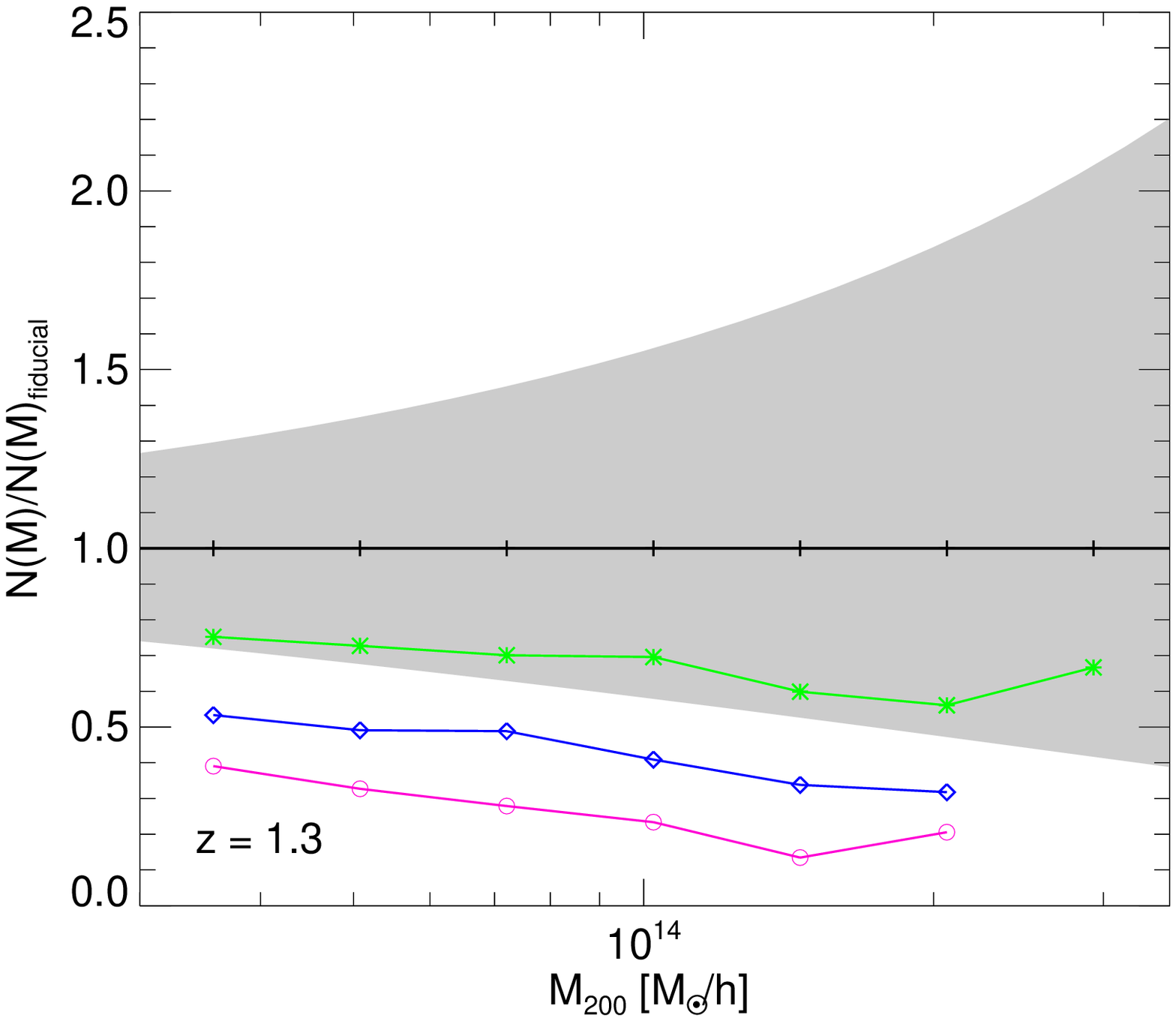}
\caption{The ratio of the { Spherical Overdensity} HMF in GR { simulations} with different neutrino masses as compared to the fiducial case of General Relativity and massless neutrinos. { Different panels refer to different redshifts as for Figs.~\ref{fig:power_GR} and \ref{fig:power_fR}. The grey shaded areas correspond to the variation of the halo mass function obtained with the standard Jenkins fitting formula \citep[][]{Jenkins_etal_2001} by varying $\sigma _{8}$ within its 2-$\sigma $ confidence interval according to the latest Planck cosmological constraints \citep[][]{Planck_016}.}}
\label{fig:massfunction_GR}
\end{figure}
\begin{figure}
\centering
\includegraphics[scale=0.45]{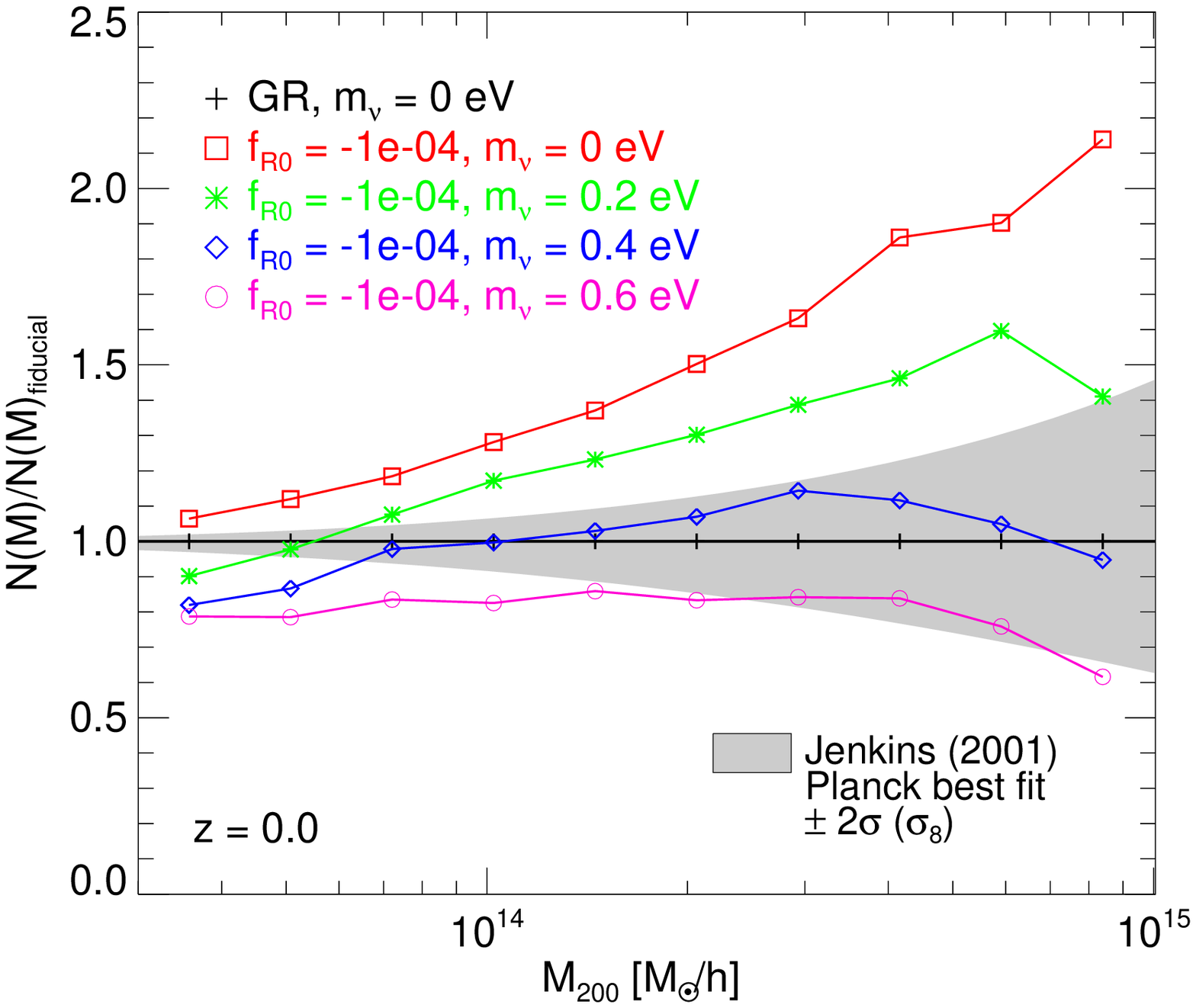}
\includegraphics[scale=0.45]{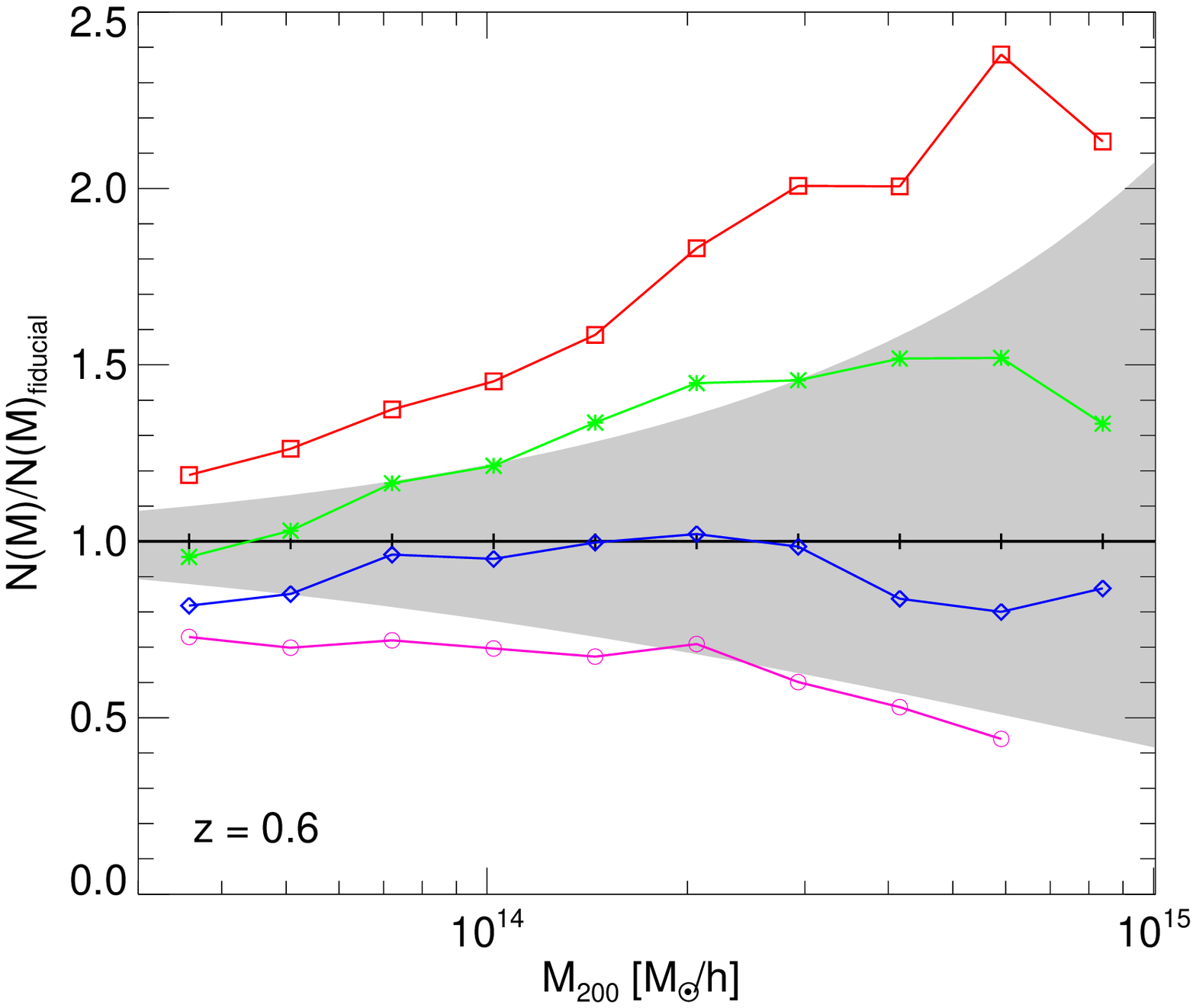}
\includegraphics[scale=0.45]{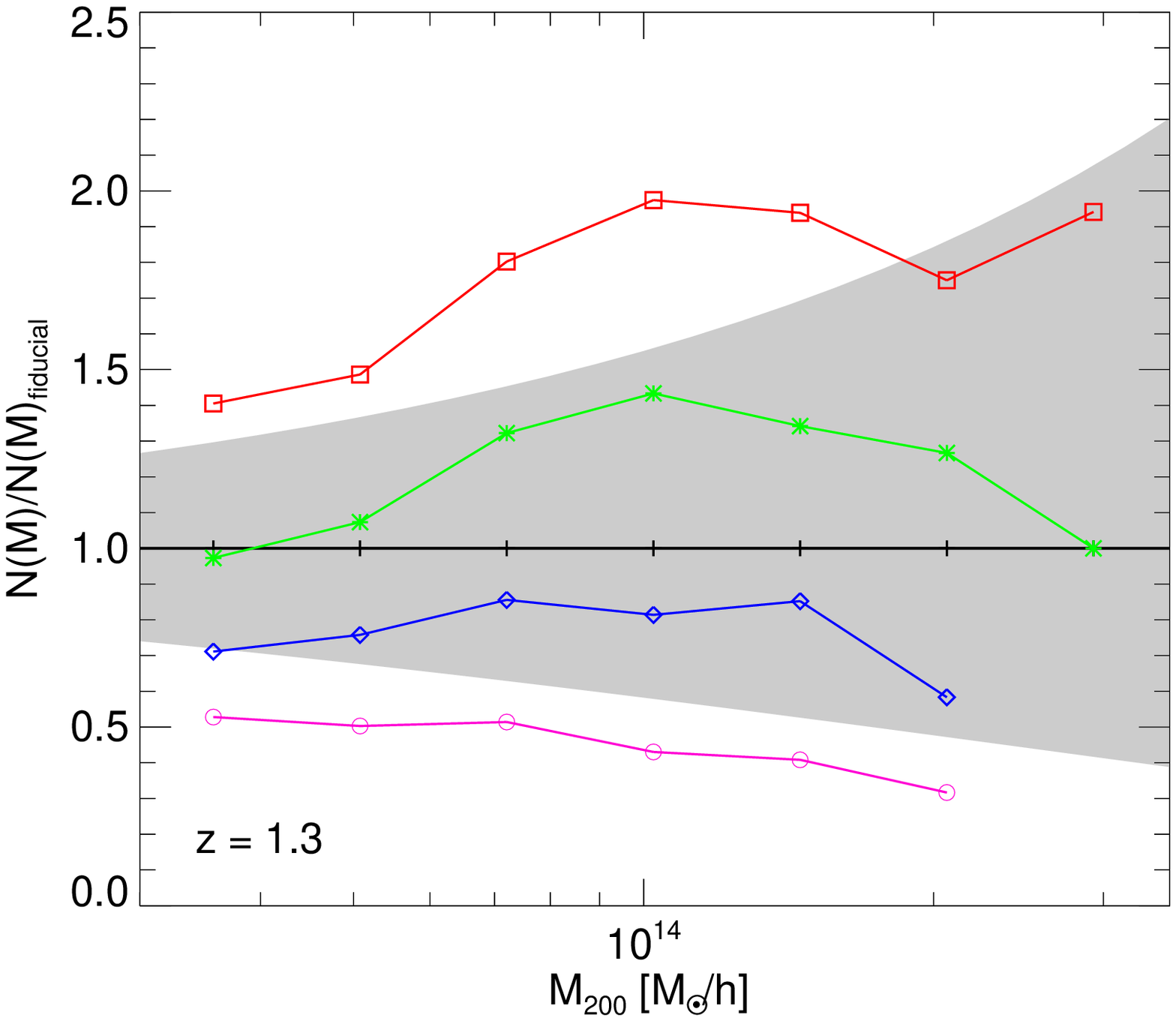}
\caption{{ As Fig.~\ref{fig:massfunction_GR} but for the combined simulations of $f(R)$ gravity and massive neutrinos.}\\
\\
\\
\\
}
%\caption{The ratio of the Spherical Overdensity HMF in $f(R)$ models with different neutrino masses as compared to the fiducial case of General Relativity and massless neutrinos at different redshifts. {The grey shaded areas correspond to the variation of the halo mass function obtained with the standard Jenkins fitting formula \citep[][]{Jenkins_etal_2001} by varying $\sigma _{8}$ within its 2$\sigma $ confidence interval according to the latest Planck cosmological constraints \citep[][]{Planck_016}.}}
\label{fig:massfunction_fR}
\end{figure}

We defer a more complete survey of other $f(R)$ cosmologies to a follow-up paper, as in this work we are mainly interested in assessing the degeneracy between $f(R)$ gravity and massive neutrinos at a semi-quantitative level. However, it is natural to expect that milder realisations of $f(R)$  theories  -- as e.g. $\bar{f}_{R0} = \left\{ -1\times 10^{-5}\,, -1\times 10^{-6}\right\}$ -- will result in a similar kind of degeneracy with lower values of the neutrino mass, and might therefore be even more difficult to disentangle observationally.

In order to compare directly the suppression effect of massive neutrinos in the two different theories of gravity, in Fig.~\ref{fig:power_comparison} we display the ratio of the matter power spectra in the massive and massless neutrino cases for GR (solid lines) and $f(R)$ (dashed lines) cosmologies, for the same redshifts displayed in Figs.~\ref{fig:power_GR} and \ref{fig:power_fR}. 
%%{ 
As one can see in the plots, the relative suppression induced by massive neutrinos is roughly the same for GR and $f(R)$ gravity, with some small deviations at the most nonlinear scales{, which can be ascribed to the different stage of evolution of the large-scale structures in which massive neutrinos evolve for the different cosmologies.} This result suggests that a sufficiently accurate estimate of the full nonlinear matter power spectrum in models that combine $f(R)$ gravity with a massive neutrino background might be obtained by applying fitting functions of the neutrino suppression obtained with standard GR simulations on top of the nonlinear matter power spectrum computed with pure $f(R)$ runs. However, such { an}
approximation should be properly tested for a wider range of neutrino masses and $f(R)$ gravity realisations before being considered sufficiently robust. 

To conclude our investigation of the matter power spectrum, we have computed the growth factor of density perturbations at three different scales $k=0.1\,, 0.5\,,1.0\, h/$Mpc by computing the square root of the ratio of the power spectrum amplitudes at the various snapshots of our simulations suite to the amplitude at $z=0$. The results of this procedure are displayed in the three panels of Fig.~\ref{fig:growth_rate}, where we plot the ratio of the growth factor of the different models under investigation to the fiducial case of GR and massless neutrinos. As one can see in the figures, GR and $f(R)$ models have (as expected) an opposite trend in the growth of density perturbations as compared to the fiducial case at all scales. Interestingly, while at linear and mildly nonlinear scales ($k=0.1\,; 0.5\, h/$Mpc)  the hierarchy of the models follows the expected linear trend at all redshifts, at the most nonlinear scales ($k=1\, h/$Mpc) the different models are found to be more entangled at low redshifts, while the growth factor slope at high redshifts shows { again the expected} linear behaviour. This { highlights} how the complex nonlinear interplay of the different gravity and neutrinos models {  produces} non-trivial effects on the evolution of density perturbations at very small scales. A detailed study of such highly nonlinear effects would require much higher resolution simulations than those available to our present suite, and we defer it to future works.
%%}

\subsection{The Halo Mass Function}

For all the simulations of our sample we have identified { dark matter} halos by means of a Friends-of-Friends (FoF) algorithm with linking length $\ell = 0.2 \times \bar{d}$, where $\bar{d}$ is the mean interparticle separation. In this case, we have considered 
only CDM particles in the linking procedure, such that the obtained halos do not include neutrino particles. 
{ This is to avoid spurious mass contamination arising from unbound neutrino particles \citep[see e.g.][]{Castorina_etal_2013, Costanzi_etal_2013}. We emphasise that the contribution of massive neutrinos to the total mass of the dark matter halo is negligible for the masses considered in this paper{, as it was recently verified with N-body simulations in the context of the standard $\Lambda $CDM cosmology \citep[][]{Villaescusa-Navarro_etal_2011, Villaescusa-Navarro_etal_2013, LoVerde_Zaldarriaga_2013}. We have directly checked that this result still holds also for $f(R)$ gravity models by comparing spherical overdensity halo masses with and without the contribution of massive neutrino particles in our $f(R)$ simulation with the largest neutrino mass (fR-nu0.6), finding deviations below $1\%$ over the whole mass range of the halo sample.}
Furthermore, for each FoF halo we have
identified its gravitationally bound substructures by means of the {\small SUBFIND} algorithm \citep[][]{Springel_etal_2001} and we associate to each FoF halo
the virial mass $M_{200}$ of its primary substructure computed as the mass of a spherical region centred { on the particle with the minimum potential} of the halo enclosing a mean overdensity $200$ times larger than the critical density of the universe. With { this} catalog of halos at hand we have computed the Halo Mass Function (HMF) by binning the halos of each cosmological simulation in $10$ logarithmically equispaced mass bins over the mass range $3.0\times 10^{13}$ M$_{\odot }/h - 1.0\times 10^{15}$ M$_{\odot }/h$, where the lower mass bound is given by the minimum halo mass resolved by the FoF algorithm for the fiducial cosmology.

In Fig.~\ref{fig:massfunction_GR} we plot for different redshifts the ratio of the Spherical Overdensity HMF obtained with {\small SUBFIND} for GR gravity with different values of the total neutrino mass to the fiducial case of GR and massless neutrinos. As expected, and consistently with previous findings {\citep[see e.g.][]{Brandbyge_etal_2010, Marulli_etal_2011, Ichiki_Takada_2012, Villaescusa-Navarro_etal_2013, Castorina_etal_2013, Costanzi_etal_2013}}, the presence of massive neutrinos significantly reduces the abundance of halos over the whole mass range allowed by our sample, with the suppression being more severe for more massive halos and for larger values of the neutrino mass. { This} effect shows a rather weak redshift dependence for $z\lesssim 1.3$. 

We now consider the combined effect of massive neutrinos and of an $f(R)$ MG theory on the HMF. In Fig.~\ref{fig:massfunction_fR} we display the ratio of the halo abundance for the $\bar{f}_{R0}=-1\times 10^{-4}$ MG model discussed in this paper with different values of the neutrino mass as compared to the fiducial cosmology with GR and massless neutrinos. As the plots show, the $f(R)$ model alone significantly enhances the abundance of halos, in particular at { large} masses. On the other hand, when MG is associated to a non-vanishing neutrino mass { this} enhancement is significantly reduced over the whole mass range available within our sample, with the largest neutrino mass included in our investigation ($\Sigma _{i}m_{\nu _{i}}=0.6$ eV) even resulting in an overall suppression of the halo abundance at all masses { and at all redshifts}.

{Analogously} to what { we} displayed in the power spectrum ratio plots above, in Figs.~\ref{fig:massfunction_GR} and \ref{fig:massfunction_fR} the grey shaded areas correspond to the mass function ratios obtained by computing the HMF through the standard Jenkins fitting formula \citep[][]{Jenkins_etal_2001} for the Planck cosmological parameters of Table~\ref{tab:parameters} but allowing $\sigma _{8}$ to vary within its 2-$\sigma $ confidence interval. { This} region  therefore { provides} a visual { indication} of the present discriminating power of observational determinations of the halo abundance at different redshifts with respect to the range of cosmological models discussed in this work.

The results show again that also for the abundance of { dark matter} halos  there is a strong degeneracy between the effects of a modification of gravity and of a non-vanishing neutrino mass, such that a suitable combination of { these} two independent extensions of the standard cosmological model { can} result in an overall abundance of { dark matter} halos hardly distinguishable from { that of} the standard fiducial $\Lambda $CDM cosmology with massless neutrinos. 

\begin{figure*}
\includegraphics[scale=0.32]{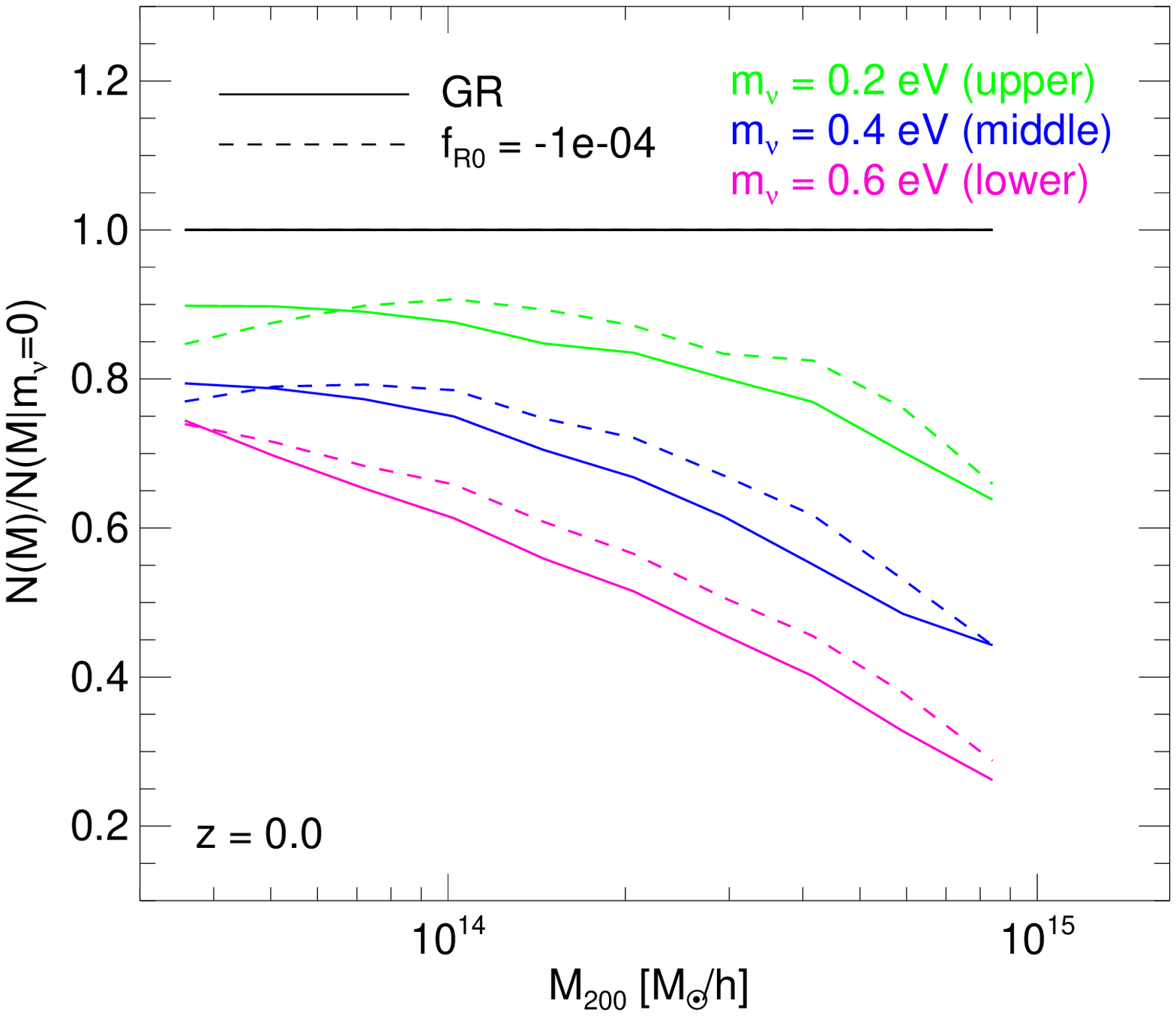}
\includegraphics[scale=0.32]{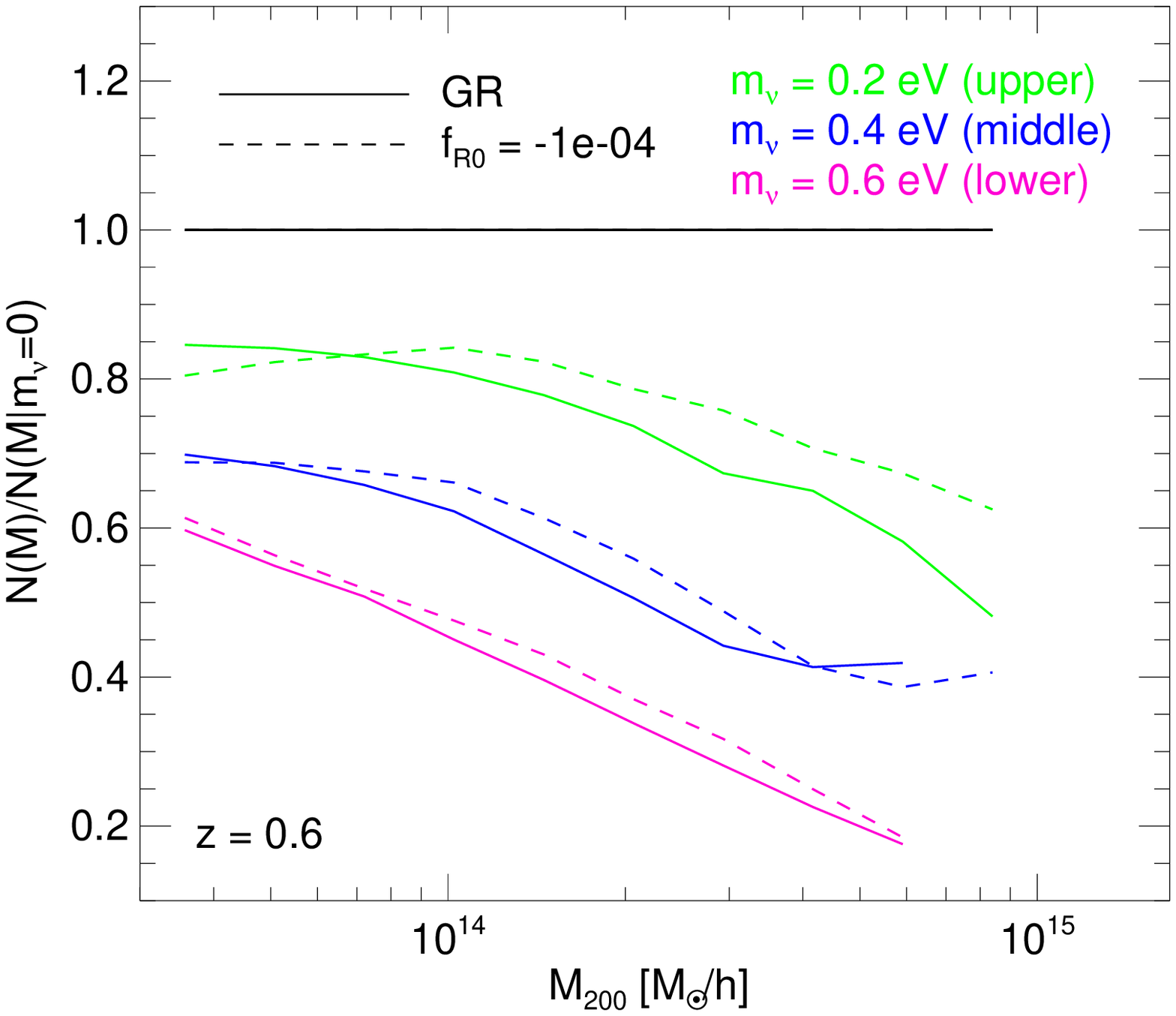}
\includegraphics[scale=0.32]{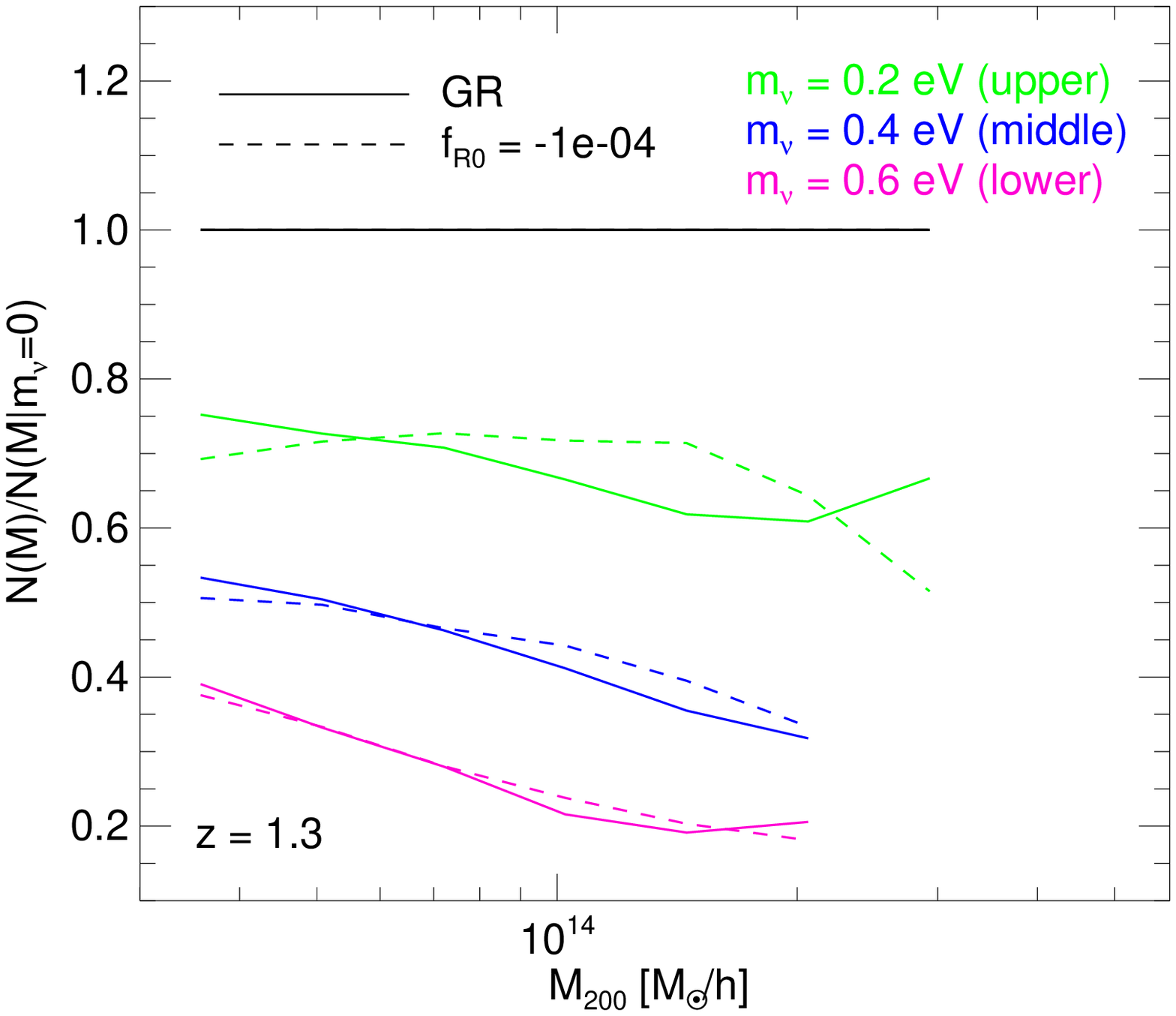}
\caption{The ratio of the Spherical Overdensity HMF for the different neutrino masses with respect to the massless case for a GR (solid lines) and $f(R)$ (dashed lines) theory of gravity { at different redshifts: $z=0$ ({\em left}), $z=0.6$ ({\em middle}) and $z=1.3$ ({\em right}).}}
\label{fig:massfunction_comparison}
\end{figure*}

\begin{figure*}
\includegraphics[scale=0.48]{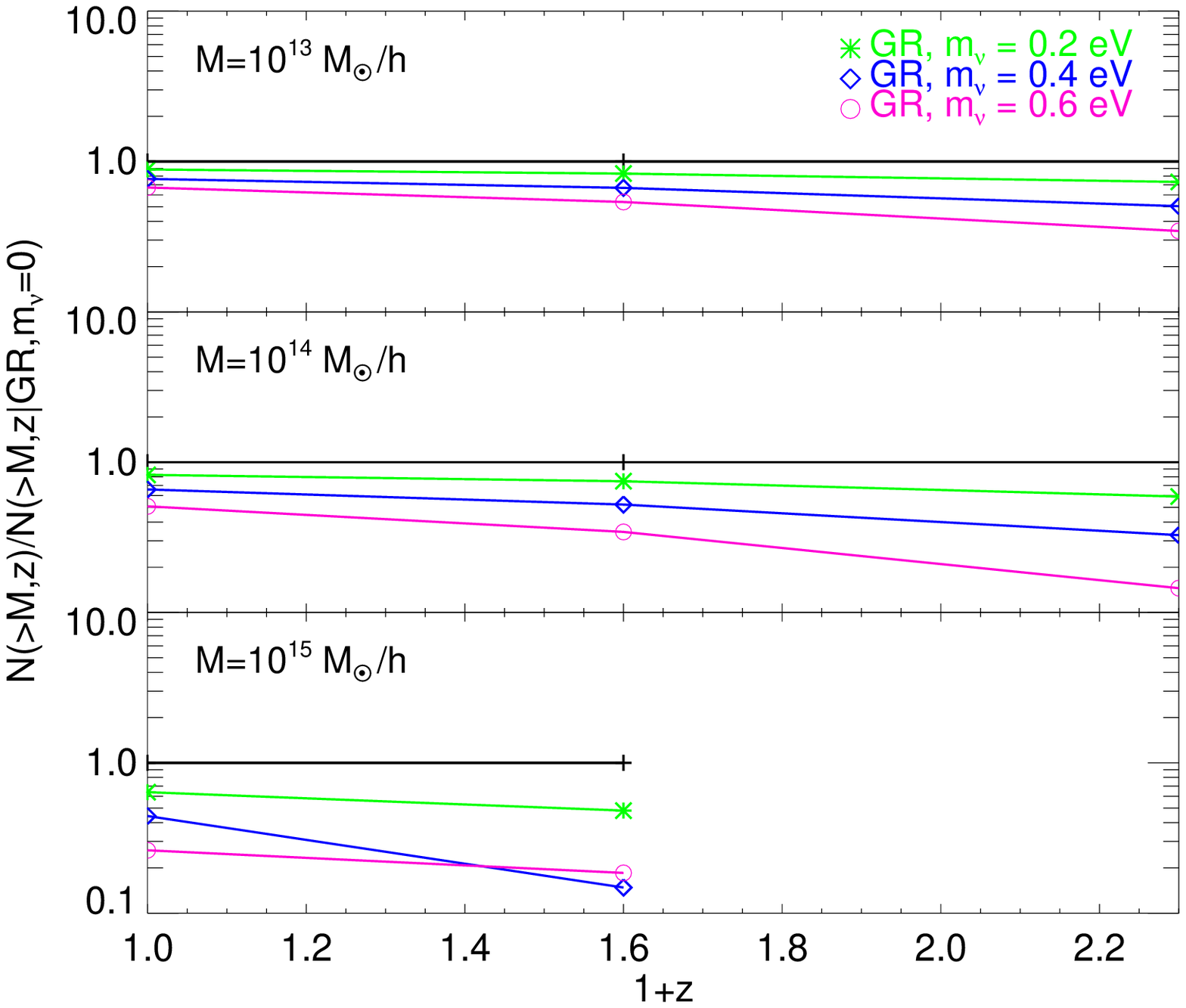}
\includegraphics[scale=0.48]{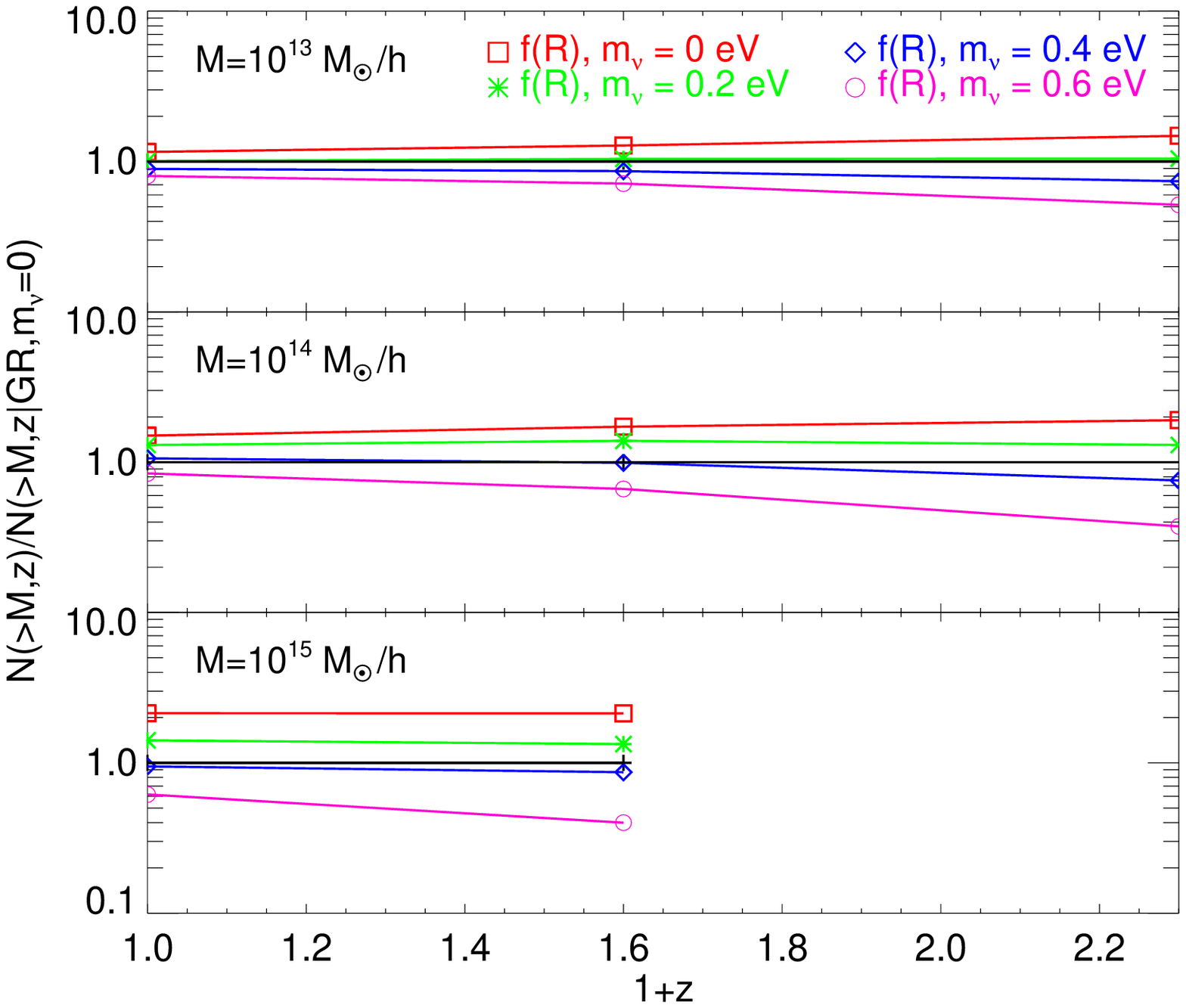}
\caption{The ratio of halo number counts above three different mass thresholds with respect to the fiducial model as a function of redshift for the GR ({\em left}) and the $f(R)$ ({\em right}) simulations with different neutrino masses.}
\label{fig:halo_counts}
\end{figure*}

In analogy to what we have done above for the nonlinear matter power spectrum, in Fig.~\ref{fig:massfunction_comparison} we show the suppression effect on the HMF of massive neutrinos of different masses as compared to the massless case in the GR (solid lines) and $f(R)$ (dashed lines) simulations. Also in this case, one can see that the neutrinos-induced suppression is roughly the same in the two theories of gravity, even though at low redshifts the $f(R)$ model systematically shows a  slightly lower suppression than the GR case. { This again appears to be related to the more evolved large-scale distribution that characterises $f(R)$ cosmologies as compared to GR at low redshifts, which makes the impact of massive neutrinos slightly less effective.}

%%{ 
Finally, in Fig.~\ref{fig:halo_counts} we display the relative abundance -- as compared to the fiducial model -- of halos above a given mass threshold $M$ as a function of redshift { and halo mass}, for the same three redshifts considered in previous figures and for three values of the threshold mass $M=\left\{ 10^{13}\,, 10^{14}\,, 10^{15}\right\}$ M$_{\odot }/h$. Therefore, this ratio shows the evolution of halo counts as a function of redshift in the different models.
From the left panel, we notice that massive neutrinos always suppress halo number counts at all masses and redshifts, while the massless neutrinos $f(R)$ model (red curve in the right panel) always shows a higher number counts than its GR counterpart. Also here, it is interesting to notice that the fiducial model lies in between the $\Sigma _{i}m_{\nu _{i}}=0.2$ eV and the $\Sigma _{i}m_{\nu _{i}}=0.4$ eV { cases} for all mass thresholds and for all redshifts, thereby confirming the degeneracy observed in all the other statistics { discussed so far}.
%%}

\subsection{Halo-matter bias}
\label{sec:bias}

\begin{figure}
\includegraphics[scale=0.45]{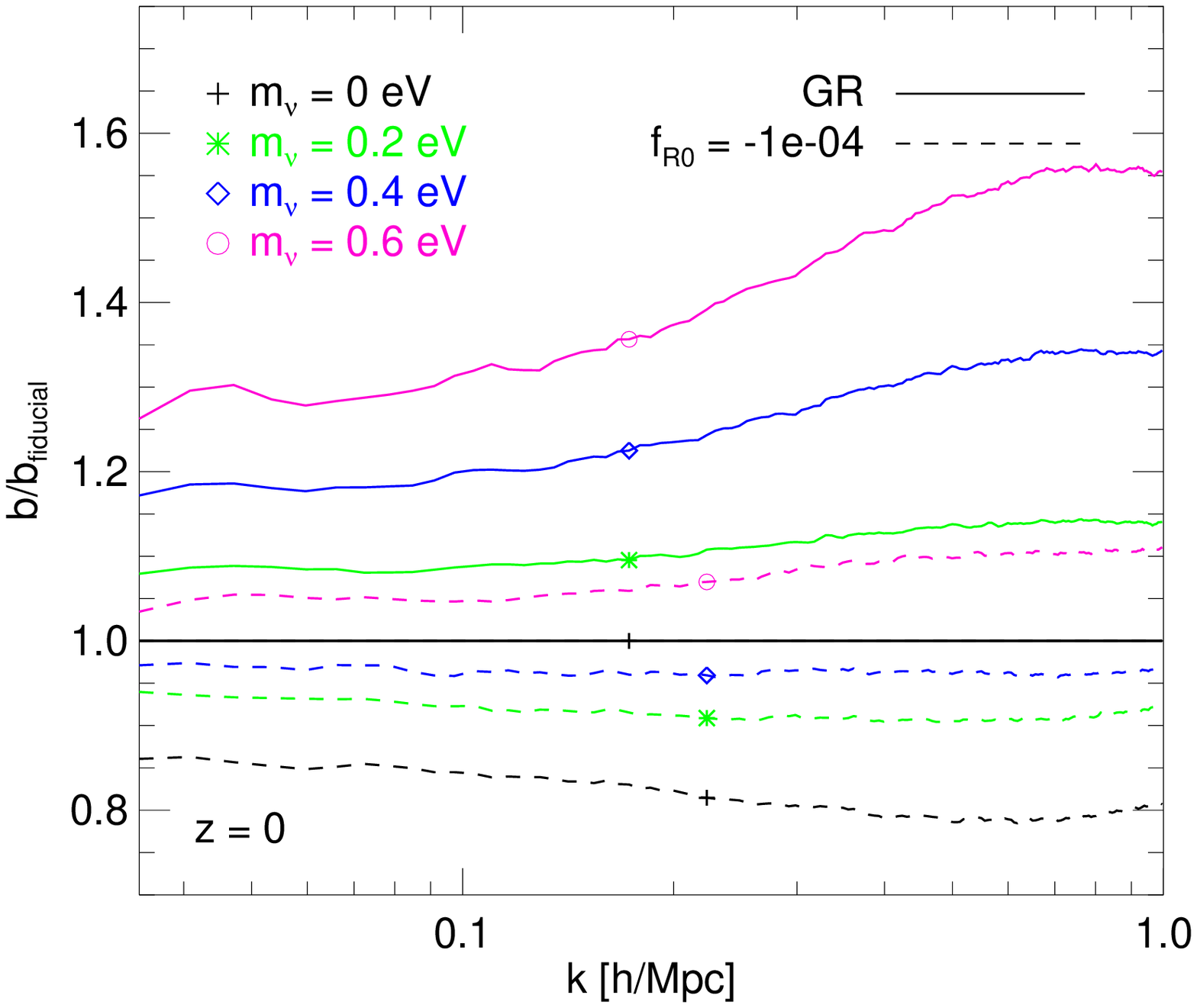}
\includegraphics[scale=0.45]{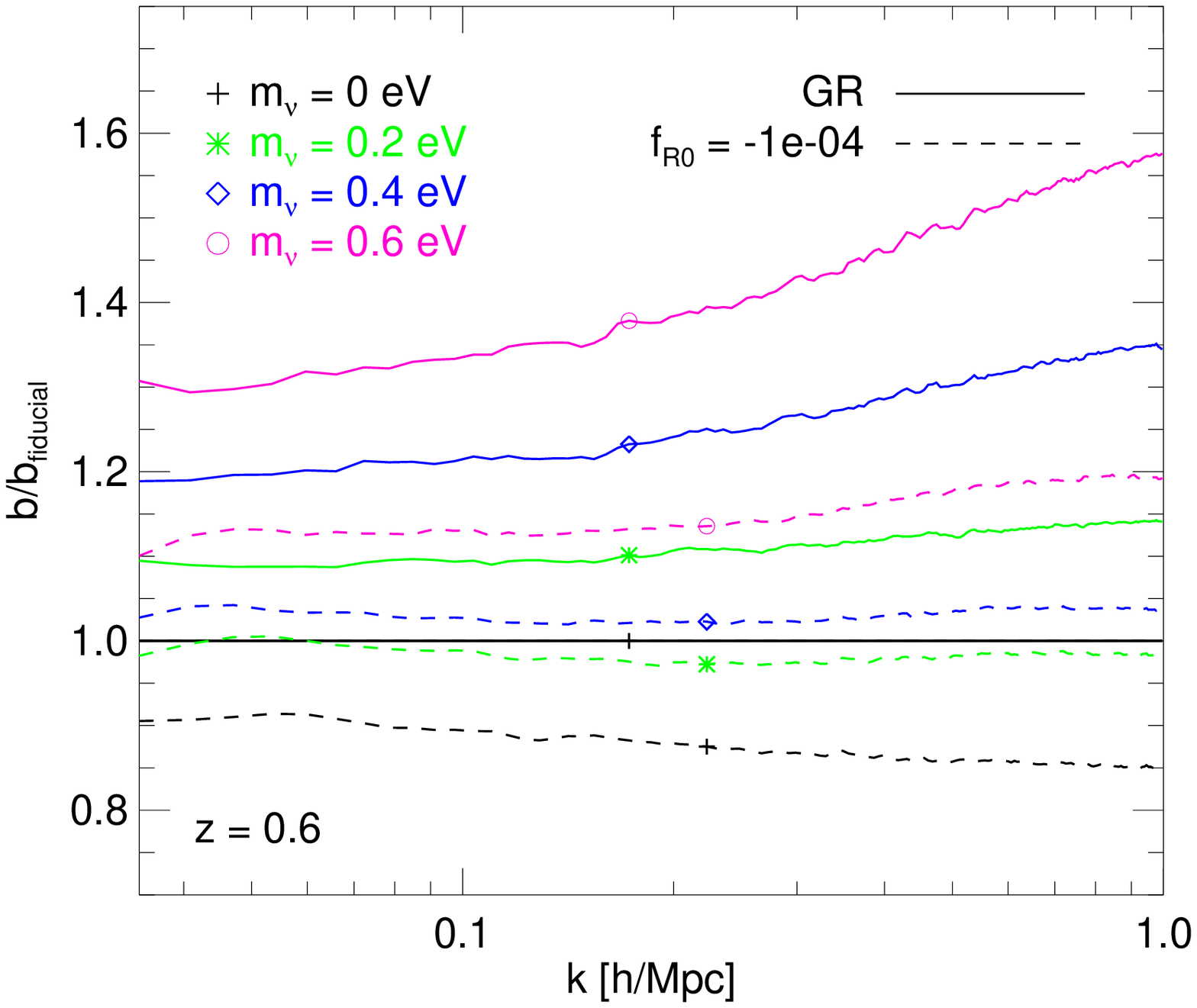}
\includegraphics[scale=0.45]{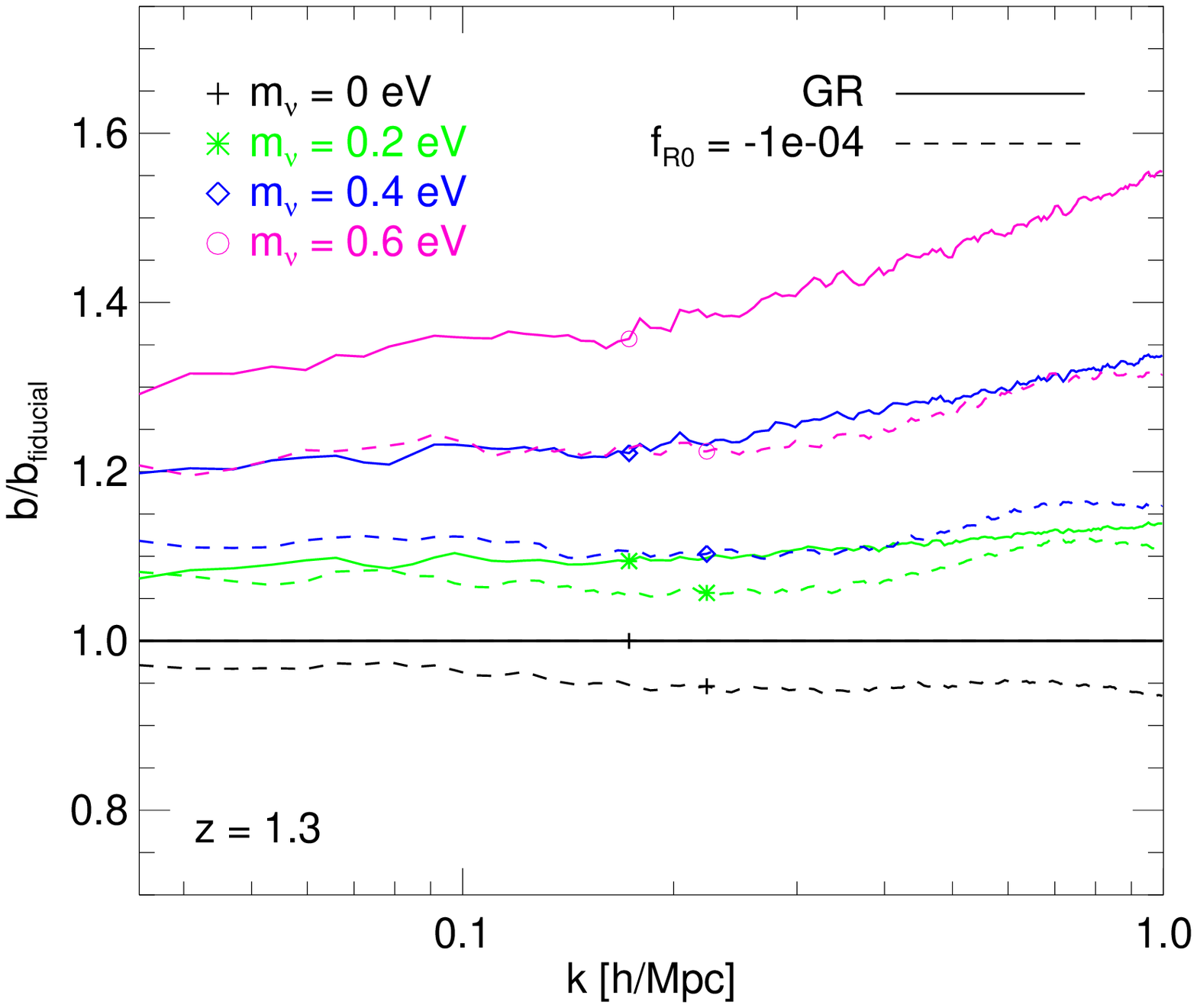}
\caption{The ratio of the halo bias for different neutrino masses in the GR and $f(R)$ cosmologies considered in the present paper (solid and dashed curves, respectively) over the bias obtained in the fiducial run with GR and massless neutrinos. { The three panels refer to different redshifts ($z=\left\{ 0\,, 0.6\,, 1.3\right\}$) as for previous figures.} }
\label{fig:bias}
\end{figure}

{ Both galaxies and dark matter halos} are biased tracers of the underlying matter distribution. In this section we focus on the bias between the spatial distribution of dark matter halos and that of the underlying 
matter for the different cosmological models considered in this paper. We compute the bias as the ratio
of the halo-matter cross-power spectrum, $P_{\rm hm}$, to the matter auto-power spectrum 
$P_{\rm mm}$
\begin{equation}
b_{\rm hm}(k)=\frac{P_{\rm hm}(k)}{P_{\rm mm}(k)}~.
\end{equation}

We have used this estimator for the bias since it does not suffer { from} shot-noise \citep[][]{Dekel_Lahav_1999, 
Hamaus_etal_2010, Baldauf_etal_2013, Smith_Scoccimarro_Sheth_2006, Baldauf_etal_2010}. Our halo catalogue consists
{ of} FoF halos with masses above $2.3\times10^{13}~h^{-1}$M$_\odot$. 
A detailed description of the method
used to compute the power spectra can be found in \cite{Villaescusa-Navarro_etal_2013b}. 

In Fig. \ref{fig:bias} we show the bias of each cosmological model normalised to the bias of the 
GR massless neutrino model. Similarly to previous figures, we show the results at $z=0$ (\textit{top}), $z=0.6$ (\textit{middle})
and $z=1.3$ (\textit{bottom}). In both GR and $f(R)$ cosmological models the effect of massive
neutrinos goes in the same direction, enhancing the bias of dark matter halos at all scales. This is due to the fact
that massive neutrinos
{
induce a suppression on the matter power spectrum which has the effect that dark matter halos of a given mass are rarer in a massive neutrino cosmology in comparison with a universe with massless neutrinos. Therefore, we expect the bias of objects of the same mass to be higher in a cosmological model with massive neutrinos.} 

In the GR cosmological model we find that massive neutrinos induce a significant scale-dependent bias 
on large scales as found in \cite{Villaescusa-Navarro_etal_2013b, Castorina_etal_2013}. This
is clearly seen in the GR $\Sigma_i m_{\nu_i}$ = 0.6 eV model (solid magenta line). In contrast,
we find that for the $f(R)$ cosmological models incorporating massive
neutrinos the large-scale bias does not exhibit any significant scale-dependence, thereby recovering the flat behaviour that characterises the standard scenario. 
{ We} will explore this 
point in further detail in a follow-up paper by using a larger suite of N-body simulations.

At $z=0$ the bias of the $f(R)$ MG model with $\Sigma_i m_{\nu_i}$ = 0.4 eV only differs from that
of the fiducial GR massless neutrino model by less than $3\%$ at all scales. At $z=0.6$ the bias
of the 0.4 eV $f(R)$ MG model only departs by less than a $5\%$ from the bias of the 
fiducial model. Larger differences among the above two models arise at $z=1.3$ ($\sim15\%$),
although these differences still do not depend on scale at any redshift.

Whereas at $z=0.6$ the bias of the $f(R)$ MG model with $\Sigma_i m_{\nu_i}$ = 0.2 eV is 
almost perfectly 
(below $2\%$) degenerate with the massless neutrinos GR cosmology, larger departures { occur}
at $z=0$ ($\sim5\%$) and $z=1.3$ ($\sim10\%$). Also in this case, we find that the differences
between the above two models does not depend on scale at any redshift.

Therefore we conclude that the degeneracies we find { for} the matter power spectrum and the 
halo mass function between $f(R)$ MG models with massive neutrinos and the massless GR 
cosmology also show up in the halo bias. { The} degeneracy is almost perfect at redshifts $z=0$ and
$z=0.6$, while it is slightly broken at $z=1.3$. Interestingly, the differences we find { for} the bias 
between the $f(R)$ MG model with $\Sigma_i m_{\nu_i}$ = 0.4 eV neutrinos and the massless 
neutrinos GR cosmology does not depend on the scale at any redshift. 

\section{Conclusions}
\label{concl}

Identifying and computing the characteristic signatures of extended cosmological models
beyond the standard $\Lambda $CDM scenario represents a necessary step in order to fully exploit the 
unprecedented accuracy of several present and future observational initiatives. On one { hand}, such { an} identification
will provide a robust interpretative framework for possible (and somewhat hoped { for}) observational { features in tension} 
with the predictions of the standard cosmological scenario; on the other { hand}, the quantitative estimation of 
the deviations expected in different observables as a consequence of specific extensions of the standard model { allows an assessment of} the effective discriminating power of present and future data sets with respect to such extended scenarios.
Given the high level of accuracy and the wide range of scales involved, { a } comparison between theoretical predictions
and observational data cannot be restricted to the background expansion history and the linear regime of structure formation, 
but needs
to be self-consistently extended to the fully nonlinear regime, thereby requiring the use of large dedicated numerical simulations. 

In recent years, a wide number of specifically-designed numerical tools have been developed with this aim, allowing to predict the impact of 
several extended cosmological models on various observables. However, { most} numerical implementations have been designed
to investigate one specific class of extended cosmologies at a time, without allowing for the possibility that the real universe
might deviate from our standard model in more than one aspect at the same time. In other words, while detailed and robust
predictions at the nonlinear level are starting to be produced for individual scenarios that extend the minimal $\Lambda $CDM cosmology 
by challenging only one among its fundamental assumptions (be it the nature of the Dark Energy, the theory of gravity, the statistical properties of the primordial density field, or the composition of the Dark Matter fraction), { these predictions usually} require that all the other standard assumptions still hold, thereby discarding possible degeneracies among different and independent extensions of the standard model.

In the present paper -- which is the first in a series of works devoted to the investigation of the possible degeneracies among different and independent extensions of the standard cosmological model -- we have { gone beyond this} oversimplified setup by including for the first time in the same numerical treatment the combined effects of a modified theory of gravity (in the form of an $f(R)$ model) and of a cosmological
background of massive neutrinos. To this end, we have combined the recent {\small MG-GADGET} implementation of $f(R)$ gravity
by \citet{Puchwein_Baldi_Springel_2013} with the massive neutrinos algorithm developed by \citet{Viel_Haehnelt_Springel_2010}, allowing us to perform the first cosmological simulations of massive neutrinos
in the context of an underlying $f(R)$ cosmology.

With such code at hand, we have run a suite of intermediate-resolution simulations of structure formation and we have investigated the
impact of { this} twofold extension of the standard model on three basic statistics of the large-scale dark matter distribution, namely the
nonlinear matter power spectrum, the halo mass function, and the halo bias. Interestingly, our results 
have revealed a 
very strong degeneracy between the observational signatures of $f(R)$ gravity and those of a massive neutrino background in all these three statistics, thereby confirming that the conservative assumption of testing different extensions of the standard model
on an individual basis is clearly too restrictive and might lead to significant misinterpretations of present and future observational data. In particular, our investigation most remarkably shows that { a} suitable combination of the characteristic parameters of $f(R)$ gravity and of the total neutrino mass, which would individually determine very significant deviations from the standard predictions in all three observables, results hardly distinguishable from the expectations of the standard cosmology. As a consequence, the discriminating power of present
and future observational constraints based on { these} statistics (or on any combination of them) with respect
to a possible modification of the laws of gravity or to a non-negligible value of the neutrino mass might be strongly weakened if { the} two
effects are both simultaneously at work. { Similarly}, null tests of the standard cosmological picture based on the same three observables
might not be { able to conclusively exclude} any of the two individual extended scenarios since their combination provides nearly indistinguishable 
predictions as compared to the reference model.\\

More specifically, we have computed the deviation with respect to the fiducial scenario (i.e. $\Lambda $CDM with massless neutrinos)
of the total nonlinear matter power spectrum for a range of simulations including both separate realisations of $f(R)$ gravity and
non-negligible neutrino masses, and their combination. As expected, the former simulations show significant deviations with respect
to the fiducial case, in full agreement with previous results in the literature. Massive neutrinos determine a scale- and redshift-dependent suppression of the matter power spectrum at both linear and nonlinear scales, with a maximum deviation from the massless case at
$k\sim 1$ {$h/$Mpc} and at $z=0$, reaching about $45\%$ for the case of a total neutrino mass $\Sigma _{i}m_{\nu _{i}} = 0.6$ eV, which is the largest value considered in the present work. On the other hand, $f(R)$ modified gravity determines an enhancement of the
nonlinear power at similar scales, with a peak of about $55\%$ at $k\sim 0.6$ {$h/$Mpc} and at $z=0$ for the specific model assumed in this work ($f_{R0}=-1\times 10^{-4}$). However, when the two effects are simultaneously included in the simulations their combined impact on the matter power spectrum is  strongly suppressed with respect to the separate individual realisations. More quantitatively, the combination 
of our $f(R)$ modified gravity cosmology with a neutrino background with total mass $\Sigma _{i}m_{\nu _{i}} = 0.4$ eV does not
deviate more than $\sim 10\%$ { in the nonlinear matter power spectrum} at $z=0$ from the fiducial model over the whole range of scales available within our simulations suite. The combined deviation is slightly larger for higher redshifts, but never exceeds $\sim 15\%$ up to $z=1.3$. These results show 
very clearly the { significant} level of degeneracy between these two independent extensions of the standard cosmological model.

Similarly, we have computed the halo mass function for all the simulations of our suite and compared the abundance of
collapsed structures in all the runs with the fiducial case of standard gravity and massless neutrinos. Again, the two separate
extensions of the standard cosmological model show very significant deviations in the abundance of structures at all
redshifts, with the massive neutrinos determining a significant suppression of the halo mass function, especially at large masses, while
our realisation of $f(R)$ modified gravity tends to produce a much larger number of halos, with an enhancement of the halo mass function up to a factor $\sim 2$
for the largest masses available in our simulations at $z=0$. Also in this case, however, when the two extensions are simultaneously
included in the numerical integration their overall combined effect is much weaker than for the separate cases just described. Interestingly,
the same combination of parameters that results the most degenerate with the fiducial model for what concerns the nonlinear matter power spectrum { is} found to provide also the most degenerate mass function, thereby preventing the possibility to use a combination of such two
observables in order to disentangle the different scenarios. 

Finally, we have performed the same type of comparison for a third basic statistics of the large-scale matter distribution: the halo-matter bias. Also in this case, both $f(R)$ modified gravity and massive neutrinos determine significant deviations in the halo bias as a function
of scale and redshift. Consistently with previous studies, we find that massive neutrinos introduce a clear scale-dependence of the halo bias as compared to the almost scale-independent behaviour of the fiducial cosmology, with larger scales showing a lower value of the bias, and with a slope of the scale-dependence that increases with the total neutrino mass. 
We find that a similar phenomenon occurs also for $f(R)$ modified gravity, although with a much weaker amplitude and an opposite trend of the scale-dependence. Again, when the two extensions of the standard cosmological model are combined in the same simulations, the
impact on the bias is significantly reduced, both in its overall amplitude and in its scale-dependence. Also in this case, at $z=0$ the most effective suppression of the predicted deviation occurs for the same combination of parameters that maximally suppress the deviation
in the other statistics. However, the combined bias seems to show a more significant redshift evolution as compared to the previous observables, thereby providing a possible way to break the (otherwise ubiquitous) degeneracy among the different cosmological models. This possibility will be further investigated in more detail in a follow-up work.\\

To conclude, we have presented in this work the first cosmological simulations that combine the effects of 
$f(R)$ modified gravity and of a cosmological background of massive neutrinos. { Our} investigation is part of a long-term programme
aimed at studying and quantifying possible degeneracies between different and mutually non-exclusive extensions of the standard
$\Lambda $CDM cosmological scenario. In this first { study}, we have demonstrated that assessing the expected features arising in different
observables as a consequence of a specific extension of the standard cosmological framework without properly taking into account 
possible degeneracies with other independent extensions might lead to significantly overestimating the constraining power of 
present and future observational data sets. In particular, we have shown that with a suitable (and not unreasonable) 
choice of parameters a combination of $f(R)$ modified gravity and of massive neutrinos might result {in a large-scale matter distribution which is} hardly distinguishable from a standard $\Lambda $CDM cosmology in several of its basic statistics.

Therefore, our results { suggest} an intrinsic theoretical limit to the effective discriminating power of present and future cosmological observations in the absence of an independent determination of the masses of neutrinos coming from particle physics experiments.

\section*{Acknowledgments}
{ We acknowledge financial contributions from contracts ASI/INAF
I/023/12/0, by the PRIN MIUR 2010-2011 ``The dark
Universe and the cosmic evolution of baryons: from current
surveys to Euclid" and by the PRIN INAF 2012 ``The Universe
in the box: multiscale simulations of cosmic structure".}
MB is supported by the  Marie Curie Intra European Fellowship
``SIDUN"  within the 7th Framework  Programme of the European Commission. 
FVN and MV are supported by the ERC Starting Grant ``cosmoIGM". 
{
EP and VS acknowledge support by the Deutsche Forschungsgemeinschaft through Transregio 33, ``The Dark Universe".
EP also acknowledges support by the ERC grant ``The Emergence of Structure during the epoch of Reionization".
The numerical simulations presented in this work have been performed 
and analysed on a number of supercomputing infrastructures: the MareNostrum3 cluster at the BSC supercomputing centre in Barcelona (through the PRACE Tier-0 grant ``SIBEL1"); the Hydra cluster at the RZG supercomputing centre in Garching; the COSMOS Consortium supercomputer within the DiRAC Facility jointly funded by STFC, the Large Facilities Capital Fund of BIS and the University of Cambridge; the Darwin Supercomputer of the University of Cambridge High Performance Computing Service (http://www.hpc.cam.ac.uk/), provided by Dell Inc. using Strategic Research Infrastructure Funding from the Higher Education Funding Council for England.}

%*****************************************************************************
\bibliographystyle{mnras}
\bibliography{baldi_bibliography}

\label{lastpage}

\end{document}